\documentclass[a4paper,11pt]{article}
\usepackage[T1]{fontenc}
\usepackage{caption}

\usepackage{color}
\usepackage{epsfig}
\usepackage{amsmath}
\usepackage{amssymb}
\usepackage{graphicx}
\usepackage{jheppub} 

\newcommand{\beq}{\begin{equation}}
\newcommand{\eeq}{\end{equation}}
\newcommand{\ft}[2]{{\textstyle\frac{#1}{#2}}}

\newcommand \re[1]{(\ref{#1})}

\def \e  {e}
\def \I  {\mathcal{W}}

\def \z  {\alpha}
\newcommand \vev [1] {\langle{#1}\rangle}

\title{\boldmath ABJM flux-tube and scattering amplitudes}

\author[\phi]{Benjamin Basso}
\author[\psi]{and Andrei V.~Belitsky}

\affiliation[\phi]{Laboratoire de Physique Th\'eorique de l'\'Ecole Normale Sup\'erieure, CNRS,\\
Universit\'e PSL, Sorbonne Universit\'es, Universit\'e Pierre et Marie Curie,\\
24 rue Lhomond, 75005 Paris, France}
\affiliation[\psi]{Department of Physics, Arizona State University, Tempe, AZ 85287-1504, USA}

\abstract{There is a number of indications that scattering amplitudes in the Aharony-Bergman-Jafferis-Maldacena theory might  have a dual description in terms of a holonomy of a supergauge connection on a null polygonal contour in a way analogous to the four-dimensional maximally supersymmetric Yang-Mills theory. However, so far its explicit implementations evaded a successful completion. The difficulty is intimately tied to the lack of the T-self-duality  of the sigma model on the string side of the gauge/string correspondence. Unscathed by the last misfortune, we initiate with this study an application of the pentagon paradigm to scattering amplitudes of the theory. With the language being democratic and nondiscriminatory to whether one considers a Wilson loop expectation value or an amplitude, the success in the application of the program points towards a possible unified observable on the field theory side. Our present consideration is focused on two-loop perturbation theory in the planar limit, begging for higher loop data in order to bootstrap current analysis to all orders in the 't Hooft coupling.}

\begin{document}

\maketitle
\flushbottom

\renewcommand{\thefootnote}{\arabic{footnote}}

\section{Introduction}\label{Sec1}

Without a doubt, integrability is a blessing in the quest of solving planar maximally supersymmetric SU$(N)$ Yang-Mills (SYM) theory in four dimensional space-time.
The gauge/string correspondence provided a hint for this profound property since it allowed one to view gauge dynamics from the perspective of a two-dimensional 
world-sheet of the type IIB string theory in the AdS$_5\times$S$^5$ target space. The existence of an infinite number of conserved charges encoding the dynamics of the two-dimensional world-sheet, and 
thus exact solvability of the string sigma model, implied its manifestation in space-time observables which are non-trivial functions of the 't Hooft coupling 
$g^2 = g_{\rm\scriptscriptstyle YM}^2 N/(4 \pi)^2$. The ones which played central roles since the inception of the AdS$_5$/CFT$_4$ correspondence were the 
scaling dimensions of composite single-trace field operators and their dual string energies; the structure constants in the Operator Product Expansion (OPE) and corresponding string couplings; last but not least, regularized gluon and open string scattering amplitudes. For this last instance, the T-self-duality of the AdS$_5\times$S$^5$ 
background was crucial since it allowed one to map the open string amplitudes to the string world-sheet bounded by a closed polygonal contour formed by the particles' 
momenta \cite{Alday:2007hr}. From the gauge theory standpoint, this yielded a conjecture that amplitudes are equivalent to the vacuum expectation value of 
a super-Wilson loop on a null polygonal contour \cite{Alday:2007hr,Drummond:2007aua,Brandhuber:2007yx,Drummond:2007cf,CaronHuot:2010ek}. By this virtue, the gauge theory enjoys yet another symmetry, the dual superconformal symmetry \cite{Drummond:2007au,Drummond:2008vq}, which is manifest in the Wilson loop representation and closes with traditional superconformal symmetry onto a Yangian algebra \cite{Drummond:2009fd}. Quantum mechanical anomalies violate the bulk of symmetries but in a manner that can be used to derive predictive Ward identities \cite{Drummond:2007au}. This allowed one to fix the four- and five-leg amplitudes completely and, starting from six legs and beyond, up to an additional dual conformal-invariant remainder function \cite{Drummond:2008aq,Bern:2005iz}.

These considerations spawned the development of a non-perturbative method to calculate the near-collinear limit of scattering amplitudes at any value of the 't Hooft coupling \cite{Alday:2010ku} by decomposing 
null-polygonal Wilson loops in terms of pentagons \cite{Basso:2013vsa}, which were determined from a set of bootstrap equations. The formalism is akin to the conventional
OPE for correlation functions of local operators. Taking the limit of adjacent segments of the loop's contour to approach the same null line generates
curvature field insertions into the Wilson link stretched along this direction. Physically, they are viewed as excitations propagating on top of the vacuum, which is the Faraday color
flux tube. Their dynamics is integrable and was explored in the context of the large-spin approximation to single-trace operators. 
At any finite order of the near-collinear expansion, there is a limited number of contributing flux-tube states, which, however, have to be summed over in order to get an exact 
representation of the Wilson loop and correspondingly space-time scattering amplitudes in generic kinematics. The pentagon program was completed in recent years \cite{Basso:2013aha,Basso:2014koa,Belitsky:2014sla,Basso:2014nra,Belitsky:2014lta,Basso:2014hfa,Basso:2015rta,Basso:2015uxa,Belitsky:2016vyq} and allowed one
to compute the aforementioned remainder function at finite coupling in the collinear limit and successfully
confront with various data stemming from other approaches to gauge-theory scattering amplitudes either within perturbation theory \cite{Goncharov:2010jf,Dixon:2011pw,Dixon:2011nj,Dixon:2013eka,Dixon:2014voa,Dixon:2014iba,Drummond:2014ffa,Dixon:2015iva,Caron-Huot:2016owq,Dixon:2016nkn} or at strong coupling \cite{Alday:2009dv,Alday:2010vh}.

A decade younger AdS$_4$/CFT$_3$ sibling of the original AdS$_5$/CFT$_4$ correspondence has been known for quite some time now. The dual pair involved in this 
case is a particular three-dimensional superconformal SU$(N)\times$SU$(N)$ Chern-Simons theory with level $\pm k$, dubbed the 
Aharony-Bergman-Jafferis-Maldacena (ABJM) theory, and M-theory on AdS$_4 \times$S$^7/\mathbb{Z}_k$. Furthermore, the double scaling limit $k, N \to \infty$ with the 't 
Hooft coupling $\lambda = N/k$ held fixed, yields a correspondence between the planar ABJM theory and free type IIA superstring theory in AdS$_4\times$CP$^3$. 

Integrability appears to be ubiquitous in both examples. However, while both instances share similarities there are also significant qualitative differences (at least in the present state of the art). The most important deviation 
from the SYM story, pertinent to our current consideration, is the absence of a well-established duality of scattering amplitudes in ABJM theory to 
a null-polygonal super-Wilson loop. This can be traced back to the lack of the fermionic T-self-duality of the AdS$_4\times$CP$^3$ background \cite{Colgain:2016gdj}, see also \cite{Adam:2010hh,Bakhmatov:2010fp,Sorokin:2011mj,OColgain:2012si}. If exists, 
it would imply by default the dual superconformal symmetry. 

In spite of the fact that this dual description is not known, the four- and six-leg tree ABJM amplitudes were found to possess a Yangian symmetry \cite{Bargheer:2010hn}. 
This can be traced back to a hidden OSp$(6|4)$ dual superconformal symmetry \cite{Huang:2010qy}. In fact, a Yangian-invariant formula for an arbitrary $n$-leg tree level amplitude 
was proposed in \cite{Lee:2010du}, see also \cite{Huang:2013owa,Elvang:2014fja}, in the form of Grassmannian integrals, mirroring the SYM construction \cite{ArkaniHamed:2009dn}. A BCFW-like recursion in three dimensions, which 
preserves the dual conformal symmetry, was suggested in \cite{Gang:2010gy}, where the eight-leg tree amplitude was calculated explicitly as well. 

Loop-level explicit ABJM analyses are more scarce, but what was found in those considerations is even more encouraging for the applicability of the pentagon OPE. The result of 
\cite{Gang:2010gy} suggested that all cut-constructible loop amplitudes within generalized unitarity-based methods \cite{Bern:1996je} possess the dual symmetry as 
well. This selection rule for the basis of unregularized momentum integrals was the central point for successful (and relatively) concise calculation of high-order perturbative 
amplitudes in the SYM theory \cite{Drummond:2006rz,Bern:2008ap}. The explicit result for the four-point ABJM planar amplitude up to two 
loops confirms this expectation. In particular, the cut-based construction of the amplitude \cite{Chen:2011vv} from a set of dual conformal invariant integrals coincides with a direct 
Feynman diagram computation \cite{Bianchi:2011dg} which does not assume this property from the onset. Moreover, the final result, in a fashion analogous to SYM, can be interpreted 
as a solution to the anomalous dual conformal Ward identities, which fix it up uniquely. This result reaffirmed the putative duality to a Wilson loop expectation value, as after proper identification it is identical to the four-cusp Wilson loop \cite{Henn:2010ps} and, in addition, is strikingly similar to its SYM counterpart. The three-loop verification was further provided in \cite{Bianchi:2014iia} as an evidence for absence of contributions to the cusp anomalous dimension in the ABJM theory at odd loop orders, also known from other considerations \cite{Gromov:2008qe}.

In ABJM theory, all multileg amplitudes beyond four external lines correspond to non-MHV ones, in the SYM language. This implies that the duality, if exists, 
should be to some version of a superloop, see e.g.~\cite{Rosso:2014oha} for a proposal. Currently, the only available higher-loop data is the six-leg amplitude which was computed at one \cite{Bianchi:2012cq,Bargheer:2012cp,Brandhuber:2012un,Brandhuber:2012wy} and two \cite{CaronHuot:2012hr} loops. It was found that its anomalous part is, again, in agreement with the results of the dual conformal anomaly equations, reproducing the BDS ansatz \cite{Bern:2005iz}. However, there is now a non-trivial homogeneous term which is the remainder function of the dual cross ratios, in complete analogy with the SYM theory.

Inspired by these observations, in this paper, we apply the pentagon paradigm to ABJM scattering amplitudes and demonstrate that, within the current state of the art, our analysis 
suggests the existence of a field theoretical observable that encodes both a (super) Wilson loop on a null polygonal contour as well as the scattering amplitudes in a single 
object. We provide some evidence for this by analyzing the OPE structure of WLs and scattering amplitudes through two loops using the pentagon factorization. Further verifications and confirmations 
require availability of higher loop perturbative data as well as multileg amplitudes.

Our subsequent presentation is organized as follows. In the next section, we briefly review the physics of the flux-tube in the ABJM theory. Some preliminary acquaintance is
expected with the subject. Next, we turn to the discussion of the pentagon transitions for all types of fundamental excitations of the flux-tube, starting with twist-one, where
our results are robust, and then turning to the twist-one-half spinons, where they are more hypothetical. We use them in Section \ref{Sec3} to construct OPEs for the bosonic
Wilson loops with six and seven points. Then, we move on to the six-leg ABJM amplitude in Section \ref{Sec4} and accommodate it within the pentagon framework. Finally, we 
discuss  problems that have to be addressed in future studies.

\section{Ans\"atze for ABJM pentagons}\label{Sec2}

In this section, we present conjectures for the pentagon transitions between flux-tube excitations in the ABJM theory. We begin with a lightning review of the flux-tube spectrum and S 
matrices, and of their relations with the $\mathcal{N}=4$ SYM flux-tube data. 
The reader is assumed to have some familiarity with the flux-tubology of $\mathcal{N}=4$ SYM. 

Prior to starting our exposition, let us point out that throughout this paper we shall use an effective coupling $g^2 = h(\lambda) = \lambda^2 + \ldots$ where $h(\lambda)$ is the 
interpolating function of the integrable spin chain of the ABJM theory. This function relates integrability to perturbation theory. It was computed at NLO in \cite{Minahan:2009aq,Leoni:2010tb} 
and is known, albeit conjecturally, to all orders in the 't Hooft coupling \cite{Gromov:2014eha}; see also \cite{McLoughlin:2008he} for its computations done at strong coupling via the string 
theory side of the dual pair. The coupling $g^2$ is also the most natural one to use for comparison between the ABJM and SYM theories.
As an illustration, the cusp anomalous dimension, which is the flux-tube vacuum energy density, can be matched between the ABJM ($\mathcal{N}=6$) and 
SYM ($\mathcal{N}=4$) theory, using integrability \cite{Gromov:2008qe}, at given coupling $g$,
\beq\label{cusps}
\Gamma_{\textrm{cusp}}^{\mathcal{N} = 4}(g) = 2\Gamma_{\textrm{cusp}}^{\mathcal{N} = 6}(g) \, .
\eeq
In particular, $\Gamma_{\textrm{cusp}}^{\mathcal{N} = 6}(g)  = 2g^2 + O(g^4)$ to leading order at weak coupling. Finally, note that since $g\sim \lambda$, the coupling $g^2$, which is the 
natural loop expansion parameter in the $\mathcal{N}=4$ theory, maps to two powers of the loop expansion parameter of the ABJM theory. The integrability formulae that we will shortly put 
forward all run in powers of $g^2$ and as such miss the odd part of physics.

\subsection{Flux tube spectrum}

Let us start by addressing the flux-tube excitations. These are effective particles which are produced when one deforms the contour of a null polygonal Wilson loop \cite{Alday:2010ku} and 
which propagate on top of the electric flux sourced by the loop \cite{Alday:2007mf}. In particular, they are produced in the collinear limit when nearby edges are set to be parallel. The 
idea behind the null Wilson loop OPE \cite{Alday:2010ku} is that a null WL can be completely flattened and replaced by multiple sums over the complete states of flux-tube excitations.

As alluded to before, flux-tube excitations can be related to field insertions along a light ray \cite{Belitsky:2011nn} or alternatively to the spectrum of large spin local operators, see \cite{Basso:2013pxa} for 
the case at hand. The latter picture allows one to obtain all-order information about their dynamics using integrability.  In particular each excitation carries a momentum $p$ for motion through
the large-spin background and an energy $E(p)$ which measures its twist. The dispersion relations $E(p)$ are known to any number of loops \cite{Basso:2010in}. The excitations were classified in 
\cite{Basso:2013pxa} and come in two types for the adjoint and bi-fundamental fields, respectively.

\begin{figure}
\begin{center}
\includegraphics[scale=0.40]{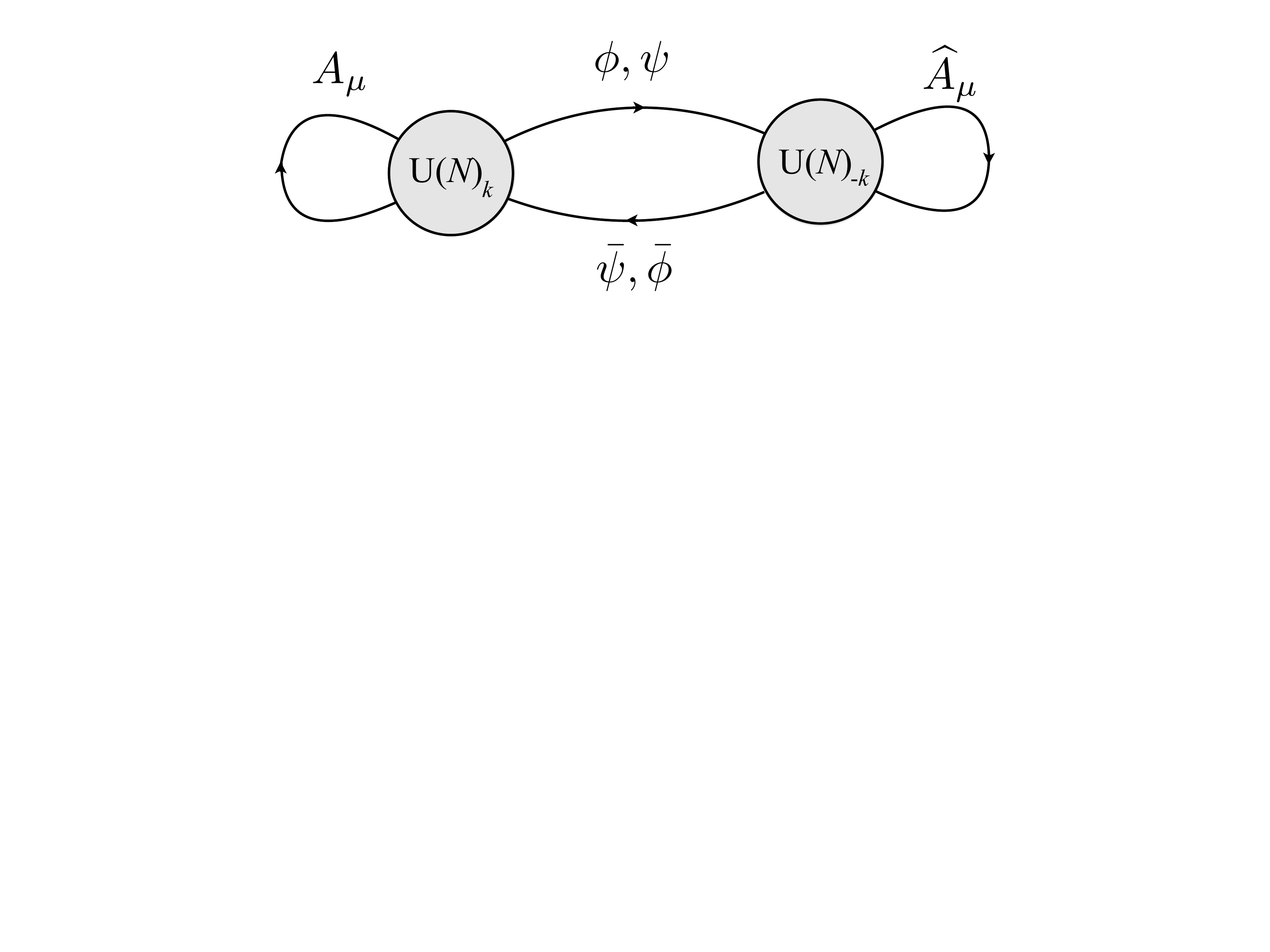}
\end{center}
\vspace{-7.5cm}
\caption{Quiver diagram for the ABJM theory.}\label{quiver}
\end{figure}

\subsubsection{Adjoints}

The adjoint excitations describe gluonic degrees of freedom and their fermionic superpartners. The most relevant bosonic excitation $F = \mathcal{F}_{11}$ corresponds to the twist 1 
component of the field strength tensor $\mathcal{F}_{\alpha\beta}$, with $\alpha, \beta$ being the spinor indices. It is the bottom representative of an infinite tower of excitations $F_{a} = 
\mathcal{D}_{11}^{a-1}\mathcal{F}_{11}$ with the twist $a = 1,2,3,...$, where $\mathcal{D}_{\alpha\beta}$ is a covariant derivative. In the integrability set up these higher twist excitations 
are not fully independent and can be seen as bound states of $a$ twist 1 gluons, $F_{a}  \sim F^{a}$.

It might be surprising to talk about gluons in a Chern-Simons-like theory where these are non-dynamical (non-propagating) degrees of freedom. We could, in principle, eliminate them using 
equations of motion and use products of bi-fundamental matter fields instead. E.g., in the large spin background, one can certainly think of the $F$ excitation as a singlet compound of 
matter fields,
\beq
F \sim \phi^{A}\bar{\phi}_{A} + \bar{\phi}_{A}\phi^{A} \, ,
\eeq
where $\phi^{A=1,2,3,4}$ denotes the scalar components of the matter hypermultiplet and $\bar{\phi}_{A}$ is its conjugate, see figure \ref{quiver}. This writing is not very useful however, if 
not for recalling the fact that whenever an $F$ appears, we should also expect a pair of matter fields as well, see e.g.~figure \ref{OPEcut}. What matters is that these compounds behave 
like single-particle excitations on the flux tube of the ABJM theory. In particular they are stable, have real dispersion relations and are to a large extent easier to deal with than the 
bi-fundamentals they are made out off at the microscopic level. They are the 3d counterparts of the gluonic modes that live on the flux tube of the $\mathcal{N}=4$ SYM theory. In the 
latter case we had two of them, $F_{a}$ and $\bar{F}_{a}$, carrying opposite charges (helicities) w.r.t.\ the transverse rotation group O(2). In the 3d theory, the transverse plane reduces 
to a line and we get a single tower of gluonic modes. Also, these 3d gluons are charge-less, since there is no (continuous) helicity group in 3d.

Up to this small departure in quantum numbers, the gluons of the 3d theory are essentially the same as those of the SYM theory. Their flux-tube dispersion relations are in fact identical to the 
ones found in the 4d theory at any coupling. In particular, one has for the twist 1 gluon,
\beq\label{EF1}
E_{F}(u)  = 1 + 2g^2 \left( \psi(\tfrac{3}{2}+iu)+\psi(\tfrac{3}{2}-iu)-2\psi(1) \right) + O(g^4)\, ,
\eeq
where $\psi(z) = \partial_{z}\log{\Gamma(z)}$ is the digamma function and where $u$ is a rapidity for the momentum of the excitation, $p_{F}(u) = 2u+O(g^2)$. Its mass starts at $1$ at weak 
coupling, since the field excitation has twist $1$, and grows up to $\sqrt{2}$ at strong coupling, where it becomes identifiable with the transverse mode of a fast-rotating string in AdS$_{4}$, 
see \cite{Frolov:2002av,Alday:2007mf,McLoughlin:2008ms,Alday:2008ut} for discussions. Notice that formula (\ref{EF1}) is 1 loop in SYM but a 2 loop result in ABJM.

The remaining adjoint particles are fermionic, $\Psi^{AB} = -\Psi^{BA}$, and fill out a vector multiplet under the R symmetry group SU$(4) \sim$ SO$(6)$, where $A, B$ are SU$(4)$ spinor indices. 
They have twist 1 and are images of the fermions of the SYM theory -- if not for the fact that in the latter theory fermions came in pairs transforming as the $\textbf{4}$ and $\bar{\textbf{4}}$ of 
SU$(4)$. The fermions cannot bind on the physical sheet and thus do not produce towers of the type we just discussed for gluons. There is something funny about them however, in a sense that they do 
have the tendency to attach to other particles at weak coupling. They then carry small momentum and minimal energy and localize on other flux tube excitations to form descendants or strings. 
The latter are not really stable, but are long lived at weak coupling and can to a large extent be viewed as particles on their own, see \cite{Basso:2014koa,Cordova:2016woh,Lam:2016rel} for 
more details. We will encounter this phenomenon latter on. For the time being, let us just add that the fermions and their funny physics is essentially identical to the one in the SYM theory. In 
particular, their dispersion relation is the same as in the 4d theory,
\beq
E_{\Psi}(u)  = 1 + 2g^2 \left( \psi(1+iu)+\psi(1-iu)-2\psi(1) \right) + O(g^4)\, ,
\eeq
with $p_{\Psi}(u) = 2u + O(g^2)$. They are (non-relativistic) Goldstone fermions for the SUSY generators that are spontaneously broken by the flux tube and, as a consequence, their mass is 1 
at any value of the coupling \cite{Alday:2007mf}.

\begin{figure}
\begin{center}
\includegraphics[scale=0.40]{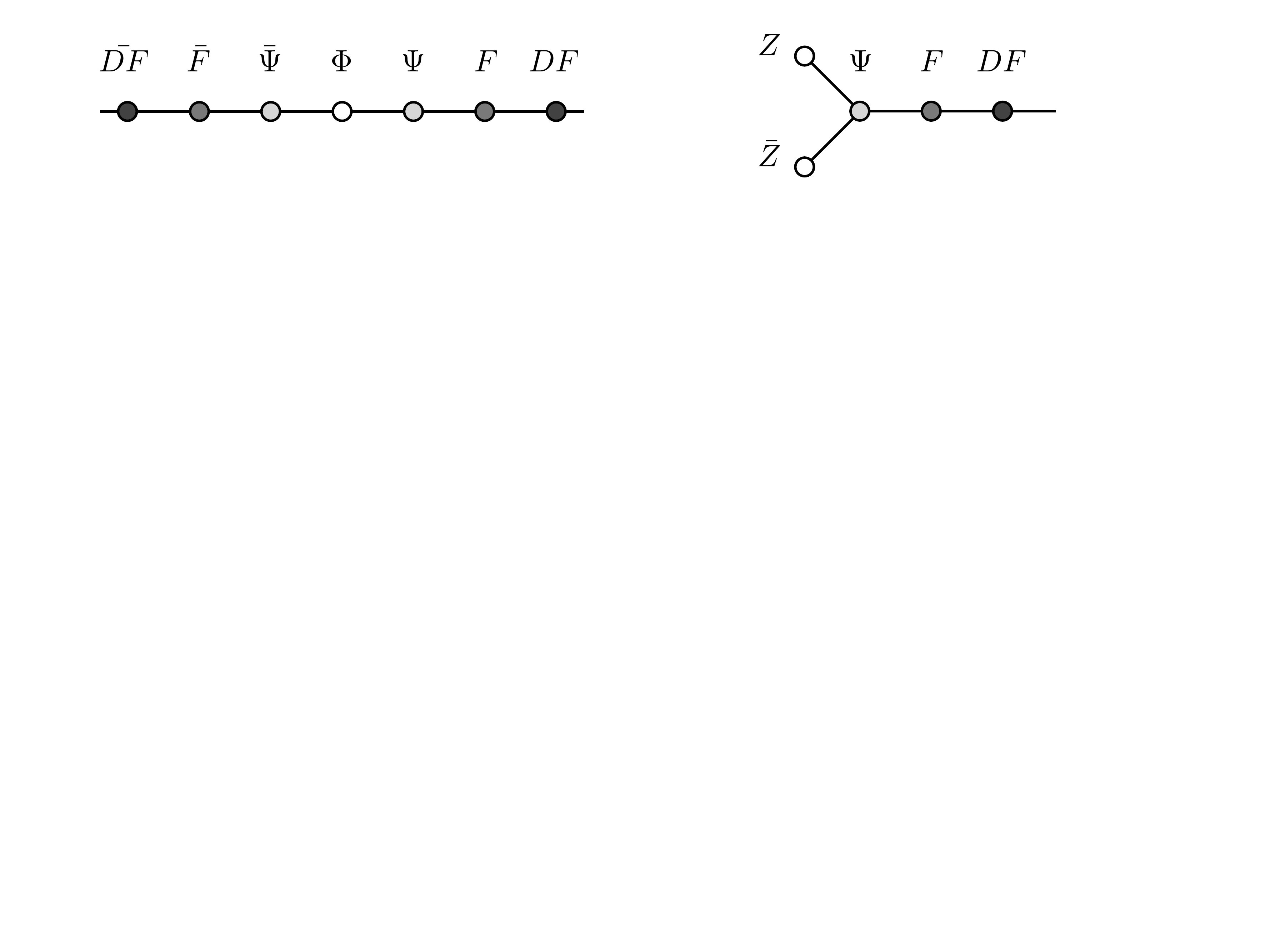}
\end{center}
\vspace{-8.5cm}
\caption{The flux tube excitations of the $\mathcal{N}=4$ and $\mathcal{N}=6$ theory can be aligned along the nodes of two infinite Dynkin diagrams of $A$ and $D$ type, respectively. The coloring 
goes along with the mass $E(p=0)$ of the excitation -- the heavier the darker. On the left panel, we have at the center the 6 scalar fields of the SYM theory surrounded by the $4+4$ twist 1 gaugino 
fields (light grey blobs). The darkest grey blobs stand for the gluonic modes: they carry no R charge but come in two infinite families of bound states (of positive and negative helicities, respectively) 
with twist $a = 1, 2, ...\,$. The right panel shows the corresponding picture for the ABJM theory. There is a single infinite tail of gluons $F_{a = 1, 2, \ldots}$ in that case.
The light grey blob on the fork is for the fermions $\Psi^{AB}$ in the $\textbf{6}$ of SU$(4)$. The lightest modes are on the fork's extremities: they are twist $1/2$ spinons $Z^{A}$ and anti-spinons 
$\bar{Z}_{A}$, in the $\textbf{4}$ and $\bar{\textbf{4}}$ of SU$(4)$, respectively.}\label{dynkins}
\end{figure}

\subsubsection{Spinons}

The remaining flux-tube excitations are bi-fundamentals. They come in conjugate pairs, $Z^{A}$ and $\bar{Z}_{A}$, called spinons and anti-spinons. They are the ABJM counterparts of the scalar 
excitations in the SYM theory and are the lightest modes on the flux tube at finite coupling. They have twist $1/2$ and belong to the $\textbf{4}$ and $\bar{\textbf{4}}$ of SU$(4)$. They carry the 
quantum numbers of the field components $(\phi^{A}|\psi_{A}; \bar{\psi}^{A}|\bar{\phi}_{A})$ of the ABJM matter hypermultiplets.  Nonetheless, they do not obviously map to either the boson or 
the fermion in these multiplets. Instead \cite{Basso:2012bw} they are solitonic excitations  -- in the sense that they interpolate between two degenerate flux tube vacua -- and they carry a fractional 
spin $1/4$. As such, we do not expect them to be easily written in terms of fundamental fields. At a coarse-grained level, they are mixtures of the two bi-fundamental fields of the ABJM theory; they 
can be produced by either field. Although a bit mysterious on the field theory side, a lot is known about them on the integrability side \cite{Bykov:2010tv,Basso:2012bw,Basso:2013pxa}. In particular, 
the energy and momentum of a spinon $Z$ with rapidity $u$ are just half of those found for a scalar $\Phi$ in the SYM theory,
\beq\label{EZ1}
E_{Z}(u)  = \tfrac{1}{2} + g^2 \left(\psi(\tfrac{1}{2}+iu)+\psi(\tfrac{1}{2}-iu)-2\psi(1)\right) + O(g^4)\, ,
\eeq
and
\beq
p_{Z}(u) = u-\pi g^2 \textrm{tanh}(\pi u) + O(g^4)\, ,
\eeq
where the $O(g^2)$ correction to the momentum is displayed for later reference.

\begin{center}
\begin{tabular}{ |p{3cm}||p{3cm}|p{3cm}| }
 \hline
 \multicolumn{3}{|c|}{Flux tube spectroscopy in 4d and 3d} \\
 \hline
type $\backslash$ theory & SYM  & ABJM \\
 \hline
 vacuum & 1 & 2 (degenerate) \\
 lightest   & $\Phi^{AB}$    &$Z^{A}$ \& $\bar{Z}_{A}$\\
 fermion & $\Psi^{A}$ \& $\bar{\Psi}_{A}$ & $\Psi^{AB}$ \\
 gluon &$F_{a}$ \& $\bar{F}_{a}$ & $F_{a}$\\
 \hline
\end{tabular}
\captionof{table}{Flux tube excitations in 4d and 3d and their correspondence.}\label{summary}
\end{center}

This is it for the content of the theory. A comparative summary of the spectra of the 3d and 4d theory is shown in table \ref{summary} and in figure \ref{dynkins}. The arrangement of flux tube 
excitations shown in figure \ref{dynkins} first appeared in \cite{Volin:2010xz} in connection with the embedding in the SYM integrable spin chain.

\subsection{Scattering matrices}

The relation between the 3d and 4d theory does not stop at the level of their energy spectra. The scattering matrices between all of these excitations are also deeply connected to one another. 
We recall these relations below. They will serve as prototypes for the pentagon transitions to be discussed shortly.

The simplest relation holds for flux tube S matrices among adjoint excitations. In this case, we have 2 excitations on the SYM side mapping to just 1 in the ABJM theory. The rule of thumb is that 
we should fold the SYM excitations to obtain the ABJM result\footnote{This procedure can be visualized by folding the Dynkin diagram of SYM  (left panel of figure \ref{dynkins})  on itself through 
the middle node.}. E.g., for the gluon S matrix, we have two 4d choices, corresponding to $FF$ and $F\bar{F}$ scattering respectively,%
\footnote{There is no backward $F\bar{F}$ scattering in this theory so $S_{F\bar{F}}$ is just a transmission phase.}
while we have only one for the ABJM theory. Hence, we write 
\beq\label{SFF}
S_{FF}(u, v)_{\mathcal{N}=6} = S_{FF}(u, v)_{\mathcal{N}=4}\times S_{F\bar{F}}(u, v)_{\mathcal{N}=4}\, .
\eeq
Higher twist gluons are bound states of $F$'s and their S matrices can be obtained by fusion. This operation commutes with the folding rule and thus the formula must also apply to them,
\beq
S_{F_{a}F_{b}}(u, v)_{\mathcal{N}=6} = S_{F_{a}F_{b}}(u, v)_{\mathcal{N}=4}\times S_{F_{a}\bar{F}_{b}}(u, v)_{\mathcal{N}=4}\, .
\eeq
The rule is more general than that since it applies to \textit{all} adjoint excitations and thus also to fermions and scattering among gluons and fermions. Fermions carry R charge indices which 
are different in the 3d and 4d theory. The matrix part of the S matrices that deal with these indices is universal and given by the SU$(4)$ rational R-matrices, in the relevant representations. 
The folding rule does not apply to them. It applies to the dynamical (a.k.a.~abelian) factors of the S matrices.

The S matrices between adjoints and spinons obey an even simpler rule since they are identical to their SYM counterparts. E.g., the S matrix between a gluon $F$ and a spinon $Z$ in the 
ABJM theory is the same as the S matrix for a gluon $F$ and a scalar $\Phi$ in the SYM theory, and more generally
\beq
S_{FZ}(u, v)_{\mathcal{N}=6} = S_{F\bar{Z}}(u, v)_{\mathcal{N}=6} = S_{F\Phi}(u, v)_{\mathcal{N}=4} = S_{\bar{F}\Phi}(u, v)_{\mathcal{N}=4}\, .
\eeq 
This sequence of equalities stays true even if $F$ is replaced by any adjoint excitation. In case where $F$ is replaced by a fermion $\Psi$ we are then referring to the dynamical factors of the 
S matrices. The rest, the actual matrix in the S matrix, are again given by SU$(4)$ R-matrices.

Last but not least, we have to discuss the pure spinon dynamics and its respective two S matrices, i.e., for the $ZZ$ and $Z\bar{Z}$ scattering. Putting aside the R-matrices, the relation to the 
SYM S matrix is now reversed since the mapping from 4d to 3d is one-to-two. We get, accordingly,
\beq
S_{ZZ}(u, v)S_{Z\bar{Z}}(u, v) = S_{\Phi\Phi}(u, v)\, ,
\eeq
where $S_{\Phi\Phi}$ is the scalar flux tube S matrix of the SYM theory. Hence, in this sector, the knowledge of the SYM S matrix is not enough to unravel $S_{ZZ}$ and $S_{Z\bar{Z}}$
individually. The missing information lies in the ratio of the S matrices,
\beq\label{su2}
S_{ZZ}(u, v)/S_{Z\bar{Z}}(u, v) = S_{SU(2)}(u-v) 
= 
\frac{\Gamma \left(\tfrac{1}{2}(iu-iv) \right)\Gamma \left(\frac{1}{2}(1+iv-iu) \right)}{\Gamma \left(\tfrac{1}{2}(iv-iu) \right)\Gamma \left(\frac{1}{2}(1+iu-iv)\right)}\, ,
\eeq
which is coupling independent and given in terms of the minimal SU$(2)$ S matrix \cite{Basso:2013pxa}.

Altogether, these relations fully characterize the flux tube S matrix of the ABJM theory in terms of the SYM one. The latter has been extensively discussed in the literature, at both weak and strong 
coupling, see e.g.~\cite{Alday:2007mf,Basso:2011rc,Dorey:2010id,Dorey:2011gr,Basso:2013aha,Basso:2013pxa,Fioravanti:2013eia,Basso:2014koa,Basso:2014nra,Bianchi:2015iza,Bianchi:2015vgw}.

\subsection{Pentagon transitions}

\begin{figure}
\begin{center}
\includegraphics[scale=0.45]{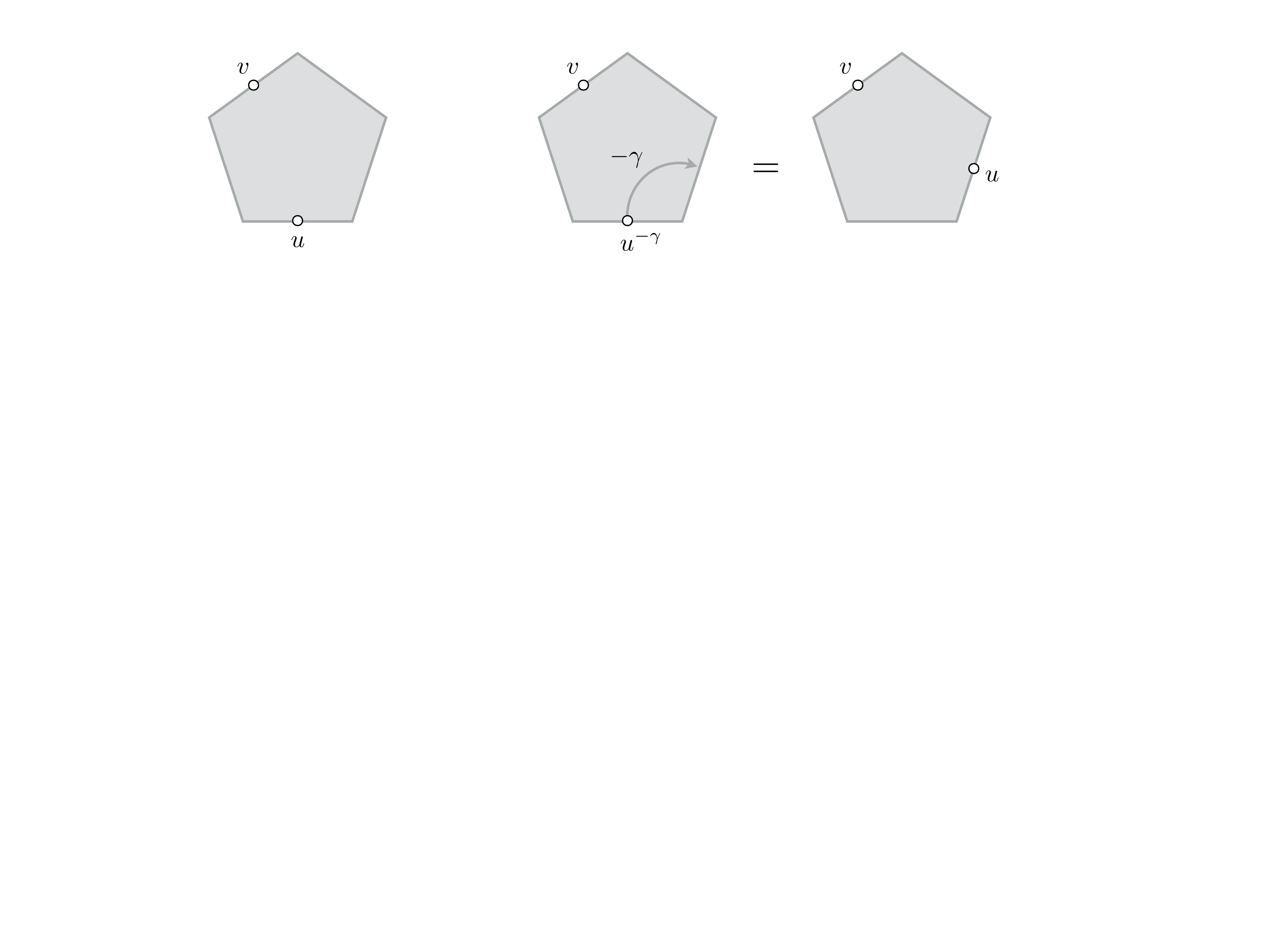}
\end{center}
\vspace{-9cm}
\caption{Left panel: cartoon for pentagon transition $P(u|v)$ between a flux tube excitation smeared with rapidity $u$ at the bottom and one with a rapidity $v$ at the top. Right panel: 
under the inverse mirror rotation $-\gamma: u\rightarrow u^{-\gamma}$ an excitation is moved anticlockwise to the neighbouring edge. The result is a pentagon transition with bottom 
and top being exchanged, $P(u^{-\gamma}|v) = P(v|u)$.}\label{pent}
\end{figure}

Next, we proceed with the pentagon transitions. These are the amplitudes for production and annihilation of excitations on the edges of a pentagon null WL \cite{Basso:2013vsa}. They are 
building blocks for the OPE decomposition of a generic null WL. The most basic pentagon transition describes a single excitation jumping from a state $|u\mathcal{i}$ to a state $|v\mathcal{i}$,
residing at the bottom and top of a pentagon, respectively, as shown in figure \ref{pent}. Their knowledge is usually enough to build all the other pentagon transitions through a factorized ansatz, see 
\cite{Basso:2013vsa,Belitsky:2014rba,Belitsky:2015efa}. In this section, we present a series of conjectures for all elementary pentagon transitions in the ABJM theory which relates them to 
their SYM counterparts, see \cite{Basso:2013vsa,Basso:2013aha,Basso:2014koa,Belitsky:2014rba,Basso:2014nra,Belitsky:2014lta} for the full list of transitions in the SYM theory and 
\cite{Basso:2015rta} for a summary. Our conjectures are robust for the adjoint excitations. The guesswork for the spinons appears to be more difficult and features a new ingredient, not present 
in the context of the SYM theory. We discuss them at the end.

\subsubsection{Pentagons for adjoints}

The most natural guess for the gluon pentagon transition in the ABJM theory is
\beq\label{PN6}
P(u|v) = P(u|v)_{\mathcal{N}=4}\times \bar{P}(u|v)_{\mathcal{N}=4}\, ,
\eeq
where $P(u|v)_{\mathcal{N}=4}$ and $\bar{P}(u|v)_{\mathcal{N}=4}$ are respectively the helicity preserving and non-preserving gluon transition of the $\mathcal{N}=4$ SYM theory. This 
conjecture has all the desired properties and verifies all the axioms imposed on the pentagon transitions. To begin with, it obeys the fundamental relation, namely
\beq
P(u|v) = S(u, v)P(v|u)\, ,
\eeq
as a result of the relations between the S-matrices of the two theories. Then, it has the right mirror property, upon the analytic continuation $-\gamma : u\rightarrow u^{-\gamma}$ of the bottom 
excitation to the neighbouring edge of the pentagon, see figure \ref{pent},
\beq
P(u^{-\gamma}|v) = P(v|u)\, .
\eeq
This property follows from the mirror properties of the SYM pentagon transitions,
\beq
P(u^{-\gamma}|v)_{\mathcal{N}=4} = \bar{P}(v|u)_{\mathcal{N}=4}\, , \qquad \bar{P}(u^{-\gamma}|v)_{\mathcal{N}=4}= P(v|u)_{\mathcal{N}=4}\, .
\eeq
It is also mirror symmetric, $P(u^{\gamma}|v^{\gamma}) = P(u|v)$, since both $P_{\mathcal{N}=4}$ and $\bar{P}_{\mathcal{N}=4}$ do possess this property. Finally, the above ansatz has a single 
pole at $v=u$, which is a kinematical singularity requirement on pentagon transitions involving identical excitations. This pole comes solely from the $P_{\mathcal{N}=4}$ factor in (\ref{PN6}). It is 
required to define the flux tube measure
\beq\label{measure-gluon}
\mu(u) = \lim\limits_{v\rightarrow u} \frac{1}{(iu-iv)P(u|v)} = \mu_{\mathcal{N}=4}(u) \times \bar{P}_{\mathcal{N}=4}(u|u) \, .
\eeq
It fixes the rule for integrating in rapidity space when considering WLs, see e.g.~Eq.~(\ref{WF}).

In the end, we could write the ansatz (\ref{PN6}) directly in terms of the S-matrix data of the $\mathcal{N}=6$ theory, with no reference to the SYM theory,
\beq
P^2(u|v) = \frac{S(u, v)}{S(u^{\gamma}, v)}\, ,
\eeq
and recognise the canonical (and the most simple form of the) ansatz for pentagon transitions. It obeys all requirements thanks to the unitarity, crossing symmetry transformation and mirror 
invariance of the gluon S matrix,
\beq
S(u, v)S(v, u) =1\, , \qquad S(u^{\gamma}, v)S(u^{-\gamma}, v) = 1\, , \qquad S(u^{\gamma}, v^{\gamma})S(v, u) = 1\, ,
\eeq
where $\gamma$ is the mirror move depicted in figure \ref{pent}.

Plugging the 4d expressions for the transitions \cite{Basso:2013aha} inside (\ref{PN6}) yields the weak coupling expression
\beq\label{PFw}
P(u|v) = -\frac{\Gamma(iu-iv)\Gamma(2+iu-iv)}{g^2\Gamma\left(-\tfrac{1}{2}+iu\right)\Gamma\left(\tfrac{3}{2}+iu\right)\Gamma\left(-\tfrac{1}{2}-iv\right)\Gamma\left(\tfrac{3}{2}-iv\right)} + O(1)\, ,
\eeq
and its residue at $iu=iv$ provides the gluon measure
\beq\label{muFw}
\mu(u) = -\frac{\pi^2 g^2}{\cosh^2{(\pi u)}} + O(g^4)\, .
\eeq
It roughly measures the cost of producing a gluon on top of the flux tube. We see that it starts at two loops, i.e.~$g^2$, in accord with the intuition that it takes a loop of matter fields to produce 
it, see figure \ref{OPEcut}.

The ansatz for the lightest gluons also determines expressions for higher twist gluons, through the fusion procedure, alluded to above, see e.g.~\cite{Basso:2014nra},
\beq\label{square}
P_{F_a|F_b}(u|v) = P_{F_{a}|F_{b}}(u|v)_{\mathcal{N}=4}\times P_{F_{a}|\bar{F}_{b}}(u|v)_{\mathcal{N}=4}\, ,
\eeq
and the associated measure as
\beq
P_{F_{a}|F_{b}}(u|v) \sim \frac{\delta_{ab}}{(iu-iv)\mu_{F_{a}}(u)}\, ,
\eeq
with $\delta_{ab}$ being the Kr\"onecker delta. To leading order at weak coupling, one finds using formulae in \cite{Basso:2014nra},
\beq
\begin{aligned}
&
P_{F_{a}|F_{b}}(u|v) 
\nonumber\\
&\quad
= 
\frac{(-1)^{b}(u^2+\tfrac{a^2}{4})(v^2+\tfrac{b^2}{4})\Gamma\left(\tfrac{a-b}{2}+iu-iv\right)\Gamma\left(\tfrac{a+b}{2}-iu+iv\right)\Gamma\left(1+\tfrac{a+b}{2}+iu-iv\right)}{g^2\Gamma^2\left(1+\tfrac{a}{2}+iu\right)\Gamma^2\left(1+\tfrac{b}{2}-iv\right)\Gamma\left(1+\tfrac{a-b}{2}-iu+iv\right)}\, ,
\end{aligned}
\eeq
while the measure takes the form
\beq\label{muFa}
\mu_{F_a}(u) = (-1)^{a} g^2\frac{\Gamma^2\left(\tfrac{a}{2}+i u\right)\Gamma^2\left(\tfrac{a}{2}-i u\right)}{\Gamma(a)\Gamma(1+a)} + O(g^4)\, .
\eeq
In distinction to what happens in SYM, these measures display infinite towers of double poles for imaginary rapidities. As we shall see later on, this feature introduces spurious singularities for the 
vacuum expectation values of the WLs. It indicates the need to have another source of contributions that will cancel them out to leading order at weak coupling, in sharp contrast 
with the SYM theory. These additions can only emerge from the spinons which we will discuss below.

The square ansatz (\ref{square}) works well for all other pentagon transitions $P_{X|Y}$ between two adjoints $X$ and $Y$, that is $F, \Psi$ and bound states $DF,$ etc. E.g., the transition 
between fermions reads
\beq\label{Pff}
P_{\Psi|\Psi}(u|v) = P_{\Psi|\Psi}(u|v)_{\mathcal{N}=4}\times P_{\Psi|\bar{\Psi}}(u|v)_{\mathcal{N}=4}\, .
\eeq
It obeys the fundamental axiom $P_{\Psi|\Psi}(u|v) = -S_{\Psi\Psi}(u, v)P_{\Psi|\Psi}(v|u)$, with the minus sign stemming for the fact that the fermion S matrix is defined such that $S_{\Psi\Psi}(u, u) = 1$.
It is harder to carry out further consistency tests since the fermions do not mirror cross nicely, see \cite{Basso:2014koa}. Nonetheless, as far as we can tell, the properties of the above ansatz 
are as good as those of the fermion proposals made for in 4d theory. Using the known expressions for the fermion transitions in the 4d theory, see e.g.~\cite{Basso:2015rta},%
\footnote{To be precise, our ansatz is $i\times (P_{\Psi|\Psi}P_{\Psi|\bar{\Psi}})$ with $P_{\Psi|\Psi}$ and $P_{\Psi|\bar{\Psi}}$ the SYM pentagons listed in \cite{Basso:2015rta}. The rescaling by an 
$i$ allows us to get a real measure $\mu_{\Psi}$.} we obtain to leading order at weak coupling,
\beq\label{PPsi}
P_{\Psi|\Psi}(u|v) = \frac{\Gamma(iu-iv)\Gamma(1+iu-iv)}{g^2\Gamma(iu)\Gamma(1+iu)\Gamma(-iv)\Gamma(1-iv)} + O(1)\, ,
\eeq
and, from the pole at $iu = iv$, we read out its measure
\beq\label{muPsi}
\mu_{\Psi}(u) = \frac{\pi^2 g^2}{\sinh^2{(\pi u)}}+ O(g^4)\, .
\eeq

The other set of transitions for which a direct lift from the 4d theory appears naturally are those involving one adjoint excitation and a spinon. These ones do not have a direct bosonic WL 
interpretation, since they do not conserve the R charge, but they are building blocks for engeneering more complicated pentagon transitions. In the SYM theory it was possible to isolate them by 
considering suitable component of the super-Wilson loop \cite{Belitsky:2014sla,Belitsky:2014lta,Basso:2015rta}. We shall not discuss this issue here as we do not know of a loop that could 
accommodate for all these excitations. (Processes with fermions might be possible to produce using the super loop of \cite{Rosso:2014oha}.)

Take as an example a mixed transition between a spinon $Z$ and a gluon $F$. A naive guess is simply that
\beq
P_{Z|F}(u|v) = P_{\Phi|F}(u|v)_{\mathcal{N}=4} = P_{\Phi|\bar{F}}(u|v)_{\mathcal{N}=4}\, .
\eeq
Here again, all axioms can be easily seen to be satisfied and self-consistent, owing in part to the fact that the RHSs are insensitive to the helicity of the adjoint excitation. We could as well replace 
$F$ by a bound state or by a fermion. In the following we will also need the pentagon transition connecting spinons and fermions and use for these the following expressions
\beq
P_{Z|\Psi}(u|v) = P_{\Phi|\Psi}(u|v)_{\mathcal{N}=4}= P_{\Phi|\bar{\Psi}}(u|v)_{\mathcal{N}=4}\, .
\eeq
We also set $P_{\bar{Z}|X}= P_{Z|X}$ for any adjoint $X$. The mixed transitions were bootstrapped on the SYM side in \cite{Belitsky:2014sla,Belitsky:2014lta}. At weak coupling, in the normalisation of \cite{Basso:2015rta}, they read
\beq\label{PFZ}
\begin{aligned}
&P_{Z|F}(u|v) = \frac{\sqrt{\frac{1}{4}+v^2}\, \Gamma(1+iu-iv)}{g\Gamma\left(\tfrac{1}{2}+iu \right)\Gamma\left(\tfrac{3}{2}-iv\right)} +O(g)\, ,\\
&P_{Z|\Psi}(u|v) = \frac{\sqrt{v}\, \Gamma\left(\tfrac{1}{2}+iu-iv\right)}{g\Gamma\left(\tfrac{1}{2}+iu\right)\Gamma(1-iv)}+O(g)\, ,
\end{aligned}
\eeq
and $P_{X|Z} = (P_{Z|X})^*$ with the involution $*$ being merely the complex conjugation. The square roots are harmless in the SYM theory; these transitions never come alone in physical 
applications and their square roots always get screened by other factors. It is less evident to 
us whether the same will always happen in the ABJM theory, but they will be of no harm in applications we consider below.

\subsubsection{Pentagons for spinons}\label{Sub:spinons}

Finally, we come to the most elaborate set of transitions, those for the bi-fundamentals. In this case, we should take a square root of sort, since the scalar field in the SYM theory maps to two 
excitations of the ABJM theory. The situation is now reversed and hence much harder. Below we present reasonable relations and assumptions for these transitions. We shall also present some 
weak coupling expressions that we will test later on.

We clearly need two pentagon transitions to characterize various processes, namely, 
\beq
P(u|v) = P_{Z|Z}(u|v)\, , \qquad \bar{P}(u|v) = P_{Z|\bar{Z}}(u|v)\, .
\eeq
It is natural, in light of the relation between the spinon and scalar excitations, to expect that
\beq
P(u|v)\bar{P}(u|v) = P(u|v)_{\mathcal{N}=4}\, ,
\eeq
where $P(u|v)_{\mathcal{N}=4}$ is the scalar transition in the SYM theory. We can therefore parameterize the spinon transitions as
\beq\label{Pansatz}
P^2(u|v) = f(u, v) \times P(u|v)_{\mathcal{N}=4}\, , \qquad \bar{P}^2(u|v) = \frac{1}{f(u, v)} \times P(u|v)_{\mathcal{N}=4}\, ,
\eeq
where $f(u, v)$ is an unknown function. We shall insist that it is such that the fundamental relation to the S-matrix is obeyed. Enforcing it, we must have
\beq\label{f/f}
f(u, v)/f(v, u) = S_{{\rm SU}(2)}(u-v)\, ,
\eeq
where the RHS is the minimal SU$(2)$ S-matrix (\ref{su2}).

However, not every solution to (\ref{f/f}) is acceptable. The function $f$ must be such that the pentagon transitions have decent singularities at weak coupling. In particular, since both $P$ and 
$P_{\mathcal{N}=4}$ have a simple pole at $u=v$, it must be so for $f$ as well,
\beq
f(u, v) \sim \frac{f'(u)}{iu-iv}\, .
\eeq
The residue $f'(u) = \partial_{u}f(u, v)|_{v = u}$ relates to the spinon measure $\mu(u) = \mu_{Z}(u) = \mu_{\bar{Z}}(u)$, canonically defined as the residue of the $P$-transition,
\beq\label{muZTomuN4}
\mu^2(u) = \frac{1}{f'(u)}\times \mu(u)_{\mathcal{N}=4}\, .
\eeq

Let us now make an educated guess for the missing ingredient, that is, the function $f$. First, recall the expression for the scalar pentagon in the SYM theory at weak coupling, which is given, 
in the normalization used in \cite{Basso:2015rta}, by
\beq\label{P4wc}
P(u|v)_{\mathcal{N}=4} = \frac{\Gamma(iu-iv)}{g\Gamma(\frac{1}{2}+iu)\Gamma(\frac{1}{2}-iv)} +O(g)\, .
\eeq
Applying the duplication formula for the Euler Gamma function,
\beq\label{dup1}
\Gamma(iu-iv) = \frac{2^{iu-iv}}{2\sqrt{\pi}}\Gamma \left(\tfrac{iu-iv}{2} \right)\Gamma \left(\tfrac{1}{2}+\tfrac{iu-iv}{2} \right)\, ,
\eeq
it can be re-written as
\beq
P(u|v)_{\mathcal{N}=4} = \frac{2^{iu-iv}\Gamma^2\left(\tfrac{iu-iv}{2}\right)}{2\sqrt{\pi}g\Gamma\left(\tfrac{1}{2}+iu\right)\Gamma\left(\tfrac{1}{2}-iv\right)}\times \frac{\Gamma\left(\tfrac{1}{2}+\tfrac{iu-iv}{2}\right)}{\Gamma\left(\tfrac{iu-iv}{2}\right)} +O(g)\, .
\eeq
This representation suggests a simple way of achieving correct analytic behavior for the ABJM transitions by choosing
\beq\label{f-weak}
f(u, v) = \frac{\z^2\Gamma\left(\tfrac{iu-iv}{2}\right)}{\sqrt{2}\Gamma\left(\tfrac{1}{2}+\tfrac{iu-iv}{2}\right)}\, .
\eeq
The choice we will make for $\z$ is to assume that it is independent of rapidities, but can in principle be a function of the coupling $g^2$. This choice fulfills the property (\ref{f/f}). Also 
the pole at $u = v$, as well as its images at $u = v+in$, are doubled, as needed to make the transition $P$ in (\ref{Pansatz}) meromorphic in $u-v$. Indeed, plugging (\ref{P4wc}) 
and (\ref{f-weak}) into (\ref{Pansatz}) yields
\beq\label{Wansatz}
\begin{aligned}
&P^2(u|v) = \frac{\z^2 2^{iu-iv}\Gamma^2\left(\tfrac{iu-iv}{2}\right)}{2\sqrt{2\pi}g \Gamma\left(\tfrac{1}{2}+iu\right)\Gamma\left(\tfrac{1}{2}-iv \right)} +O(1) \, , \\
&\bar{P}^2(u|v) = \frac{2^{iu-iv}\Gamma^2\left(\tfrac{1}{2}+\tfrac{iu-iv}{2}\right)}{\z^2 \sqrt{2\pi} g \Gamma\left(\tfrac{1}{2}+iu\right)\Gamma\left(\tfrac{1}{2}-iv\right)} + O(1) \, .
\end{aligned}
\eeq
From the first line, we also read out the measure 
\beq\label{Mansatz}
\mu^2(u) = \frac{\pi\sqrt{\pi}g}{\z^2\sqrt{2}\cosh{(\pi u)}} + o(g)\, .
\eeq
Equations (\ref{Wansatz}) and (\ref{Mansatz}) are the expressions that we will put to test later on. (In particular, comparison with the two loop hexagon WL will enforce that $\alpha^4 = 1+O(g)$.)

\begin{figure}
\begin{center}
\includegraphics[scale=0.45]{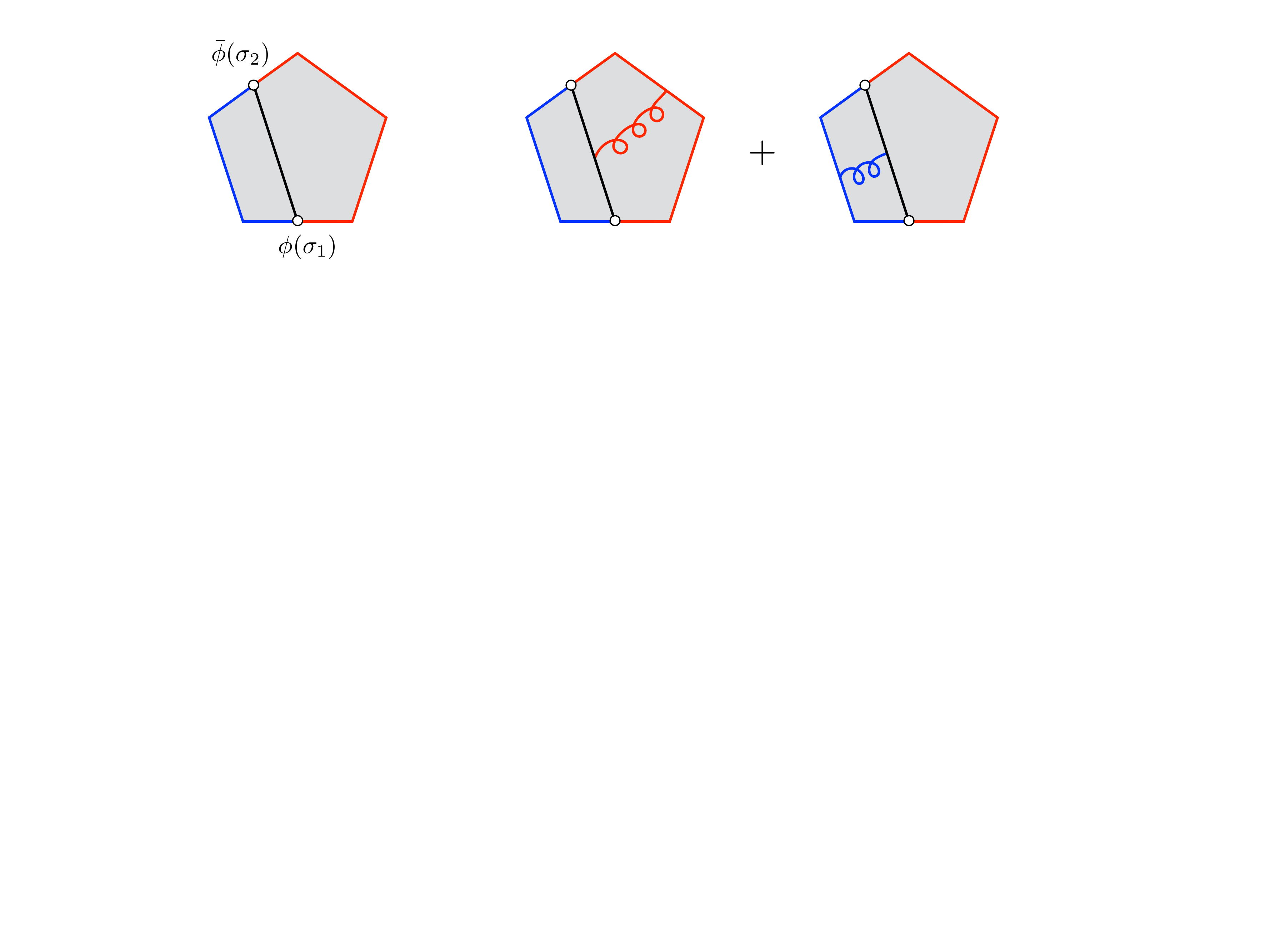}
\end{center}
\vspace{-9cm}
\caption{Tree level representation of the pentagon transition for matter fields, here taken to be scalars, and its first loop corrections. The two colors (red and blue) refer to the two gauge fields of 
the ABJM theory. Analyzing the first loop corrections could help understanding how to upgrade pentagon transitions to higher orders.}\label{prop}
\end{figure}

The choice (\ref{f-weak}) appears quite natural at weak coupling where the transition $P(u|v)$ should relate to tree level propagators for matter fields inserted along the pentagon WL; see 
figure \ref{prop} for an illustration. The square roots in the transition appear worrisome in this regard. There is however no obvious relation between the implicit normalization implied by 
our ans\"atze and the one needed to represent the direct tree-level insertions of fields along the loop. In other words, we can assume that the pentagon transition (\ref{Wansatz}) describes 
a free propagator for suitably smeared insertions.

After stripping out conformal weights of the scalar field, see \cite{Basso:2013aha,Belitsky:2014rba} for a detailed discussion, its propagator reads
\beq\label{phi-prop}
\mathcal{h}\phi(\sigma_{1})\bar{\phi}(\sigma_{2})\mathcal{i} = \frac{1}{\sqrt{e^{\sigma_{1}-\sigma_{2}} + e^{\sigma_{2}-\sigma_{1}}+e^{\sigma_{1}+\sigma_{2}}}} = \int \frac{du dv}{(2\pi)^2} e^{-iu \sigma_{1}+iv\sigma_{2}}P_{\phi|\phi}(u-i0|v)\, ,
\eeq
where
\beq
P_{\phi|\phi}(u|v) =  \Gamma\left(\tfrac{1}{4}-\tfrac{iu}{2}\right)\Gamma\left(\tfrac{iu-iv}{2}\right)\Gamma\left(\tfrac{1}{4}+\tfrac{iv}{2}\right)\, .
\eeq
Similarly, for the twist 1/2 component of the fermion $\psi$, we get
\beq
P_{\psi|\psi}(u|v) =  \Gamma\left(\tfrac{3}{4}-\tfrac{iu}{2}\right)\Gamma\left(\tfrac{iu-iv}{2}\right)\Gamma\left(\tfrac{3}{4}+\tfrac{iv}{2}\right)\, .
\eeq
Both relations follow from the following general formula
\beq\label{formula-s}
\int\frac{du dv}{(2\pi)^2} \Gamma\left(s-\tfrac{iu}{2}\right)\Gamma\left(\tfrac{iu-iv+0}{2}\right)\Gamma\left(s+\tfrac{iv}{2}\right) e^{-iu\sigma_{1}+iv\sigma_{2}} 
= 
\frac{4\Gamma(2s)}{(e^{\sigma_{1}-\sigma_{2}}+e^{\sigma_{2}-\sigma_{1}}+e^{\sigma_{1}+\sigma_{2}})^{2s}}\, ,
\eeq
used above for the conformal spins $s = 1/4$ and $s = 3/4$ for $\phi$ and $\psi$ fields, respectively. Now, clearly, one can find smearing factors for the incoming and outgoing flux tube 
states such that the transition $P$, dressed with the measures, satisfies
\beq
\mathcal{N}_{\phi}(u)\mu(u)P(u|v)\mu(v)\mathcal{N}^*_{\phi}(v) \propto P_{\phi|\phi}(u|v)\, ,
\eeq
up to an irrelevant overall factor, and similarly for $P_{\psi|\psi}$. For instance, we can choose
\beq
\mathcal{N}_{\phi}^2(u) \propto \frac{\Gamma\left(\tfrac{1}{4}-\tfrac{iu}{2}\right)}{\Gamma\left(\tfrac{3}{4}-\tfrac{iu}{2}\right)}\, ,
\eeq
for the smearing factor relating the scalar insertion to our abstract spinon, and
\beq
\mathcal{N}_{\psi}^2(u) = 1/\mathcal{N}_{\phi}^2(u)\, ,
\eeq
the one of the fermion. We will re-encounter these smearing factors later on in the flux tube analysis of scattering amplitudes, although combined differently. Smearing factors also 
showed up in the SYM theory in the study of non MHV amplitudes \cite{Belitsky:2015efa} and were dubbed non MHV form factors \cite{Basso:2013aha}. Their structure was simpler 
and easier to understand thanks to their relation to supersymmetry generators. We do not understand them that well in the current 3d story. It is therefore difficult to make precise 
the mapping between the integrability based predictions and field theory WLs with insertions at higher loops. However one might be able to learn about the higher loop structure of 
the pentagon transitions by considering dressed propagators like the one depicted in the right panel of figure \ref{prop}.

In light of this agreement, it is tempting to lift the ansatz (\ref{f-weak}) to an all-order conjecture. Equation (\ref{f/f}) for $f$ is coupling independent and function of the difference of rapidities 
only. It is then natural to look for a solution possessing the same properties.

There is a problem however with the mirror axiom. Namely, the function (\ref{f-weak}) transforms badly upon the mirror rotation and the weak coupling singularities that it is removing on one 
sheet eventually re-emerge on its mirror rotated version. The SYM transition itself is mirror symmetric,
\beq
P_{\Phi|\Phi}(u^{-\gamma}|v) = P_{\Phi|\Phi}(v|u)\, .
\eeq
The problem comes from the function $f$. The inverse mirror rotation $-\gamma : u\rightarrow u^{-\gamma}$ boils down to a shift by $-i$ on any meromorphic function, but $f$ does not 
map back to itself under this shift,
\beq
f(u^{-\gamma}, v) = f(u-i, v) = -i\, \textrm{tanh}{(\pi(u-v))}\times f(v, u) \neq f(v, u)\, .
\eeq
Therefore, it is hard to believe that $f$ will remain the same at any loop order.

We could enlarge our ansatz by promoting $\alpha$ in (\ref{f-weak}) to a symmetric function of rapidities $\alpha \rightarrow \alpha(u, v) = \alpha(v, u)$ and look for a solution with a cut structure 
permitting both $\alpha \sim 1$ at weak coupling and $\alpha^2(u^{-\gamma}, v)  = i\,\textrm{coth}{(\pi(u-v))}\alpha^2(v, u)$ at finite coupling. The space of solutions is huge and we do not 
even know if this factor admits an expansion in integer powers of $g^2$, like everything else so far, or if odd loops should be included as well. Odd loop corrections to null polygonal Wilson loops are 
not excluded; although they were found to cancel out at one loop \cite{Henn:2010ps,Bianchi:2011rn,Wiegandt:2011zz}. If they exist and if our other conjectures are correct, then they must necessarily sit inside 
the function $\alpha$. (Progress with this issue might be accessible without necessarily computing loop corrections to higher polygonal WLs. Investigation of the loop corrections to the pentagon 
WL showed in figure \ref{prop} should already provide some insights into the structure of the extra term in $P$.) In lack of information on the class of functions we are after, it appears difficult, if 
not impossible, to pin down the right solution for $\alpha$.%
\footnote{Another source of inspiration for this problem is strong coupling where the pentagon transitions should map to form factors of twist operators in Bykov model \cite{Bykov:2010tv}. There 
are several candidates here again and we could not find a single hint as to how to solve the problem all the way down to weak coupling.}
In this paper we shall stick to our naive ansatz and treat $\alpha$ as a constant. Although it is unlikely to be valid at higher loops, it will be sufficient for the weak coupling data that we shall 
analyze in subsequent sections.

Let us add in conclusion that it is possible to find a simple function $f$ that obeys both the fundamental relation and the mirror axiom. For instance,  
\beq
\alpha^2(u, v) \propto \textrm{sech}{(\pi(u-v))} \qquad \Rightarrow \qquad f(u, v) \propto \Gamma\left(\tfrac{iu-iv}{2}\right)\Gamma\left(\tfrac{1}{2}-\tfrac{iu-iv}{2}\right)
\eeq
does obey both of them. This choice is natural at strong coupling and relates to the minimal form factors for twist operators in Bykov model \cite{Bykov:2010tv}. However, since it is not a perfect square, it yields 
unwieldy singularities at weak coupling and as such does not appear as a viable option.

\section{Wilson loops}\label{Sec3}

Equipped with a set of pentagon transitions, we can move on to the actual computation of the null polygonal Wilson loop in the ABJM theory. The latter is defined in the usual fashion as a 
vacuum expectation value of a path ordered exponential of a gauge field integrated along a contour $C_{n}$,
\beq\label{Wn}
W_{n} = \frac{1}{N} \left\langle \textrm{tr}\, P e^{i\int_{C_n} dx \cdot A } \right\rangle = W_{n}^{\rm BDS}\times \mathcal{R}_{n}\, .
\eeq
Here $C_{n}$ describes a null polygon with $n$ edges and $A$ can be either of the two gauge fields of the ABJM theory, see figure \ref{quiver}. In this paper we shall remain agnostic about 
which gauge field is running around the loop. To the accuracy that we will be working, there is simply no difference between the two options \cite{Henn:2010ps,Bianchi:2011rn,Wiegandt:2011uu}. (The difference 
is odd in the coupling and stays beyond the range of applicability of our conjectures; it could contain important information about higher loop completion of our ans\"atze, however).

In Eq.\ (\ref{Wn}), we anticipated a factorization of the Wilson loop into a BDS part and a remainder function, with the former absorbing all the UV divergences and the latter being a finite function 
of conformal cross ratios. This decomposition, which is a consequence of the dual conformal Ward identities in the SYM theory \cite{Drummond:2007au}, was also observed to be true perturbatively 
in the 3d theory \cite{Henn:2010ps,Bianchi:2011rn,Wiegandt:2011uu}. Moreover, and quite remarkably, the remainder function $\mathcal{R}_{n}$ vanishes through two loops for all polygons 
\cite{Henn:2010ps,Bianchi:2011rn,Wiegandt:2011uu},
\beq
\mathcal{R}_{n} = 1+O(g^3)\, ,
\eeq
meaning that WLs in the ABJM and SYM theory are the same to leading order at weak coupling, if not for the difference in the cusp anomalous dimension, see (\ref{cusps}).

In this section, we will apply our formulae to the computation of the two-loop hexagonal and heptagonal loops for the lowest two twists in the multi-collinear limit, reproducing  available 
perturbative results. We shall also provide a prediction for logarithmically enhanced terms, or leading OPE discontinuity, of the hexagon loop at four loops. Finally, we shall subject our 
conjectures to a test at strong coupling, by comparing them with the leading twist corrections to the areas of minimal surfaces in AdS$_{4}$.

Our analysis relies on the previously derived expressions for the pentagon transitions. We also assume that the multi-particle integrands take the usual form and factorize into products of pentagon 
transitions \cite{Belitsky:2014rba,Basso:2014nra,Belitsky:2014lta}, for the dynamical parts, and rational functions of rapidities \cite{Basso:2013aha,Basso:2014jfa,Belitsky:2016vyq}, for the 
matrix parts. More specifically, specializing to the hexagon WL for simplicity, we assume that the OPE integrand for a flux tube state made out of $n$ excitations $A_{i}(u_{i})$, with $i=1, \ldots , n$, 
takes the form
\beq
\frac{\prod_{i}\mu_{A_{i}}(u_{i})}{\prod_{i\neq j}P_{A_{i}|A_{j}}(u_{i}|u_{j})}\times \Pi(\{u_{i}\})\, ,
\eeq
where $\Pi(\{u_{i}\})$ is the matrix part. The latter can be obtained using an integral formula \cite{Basso:2014jfa} or by contracting the matrix pentagons of \cite{Belitsky:2016vyq}. We cannot 
confidently predict the sign of each contribution however. These signs will be fixed through a comparison with perturbative results -- and more specifically through the condition that spurious 
singularities cancel out globally.

\begin{figure}
\begin{center}
\includegraphics[scale=0.30]{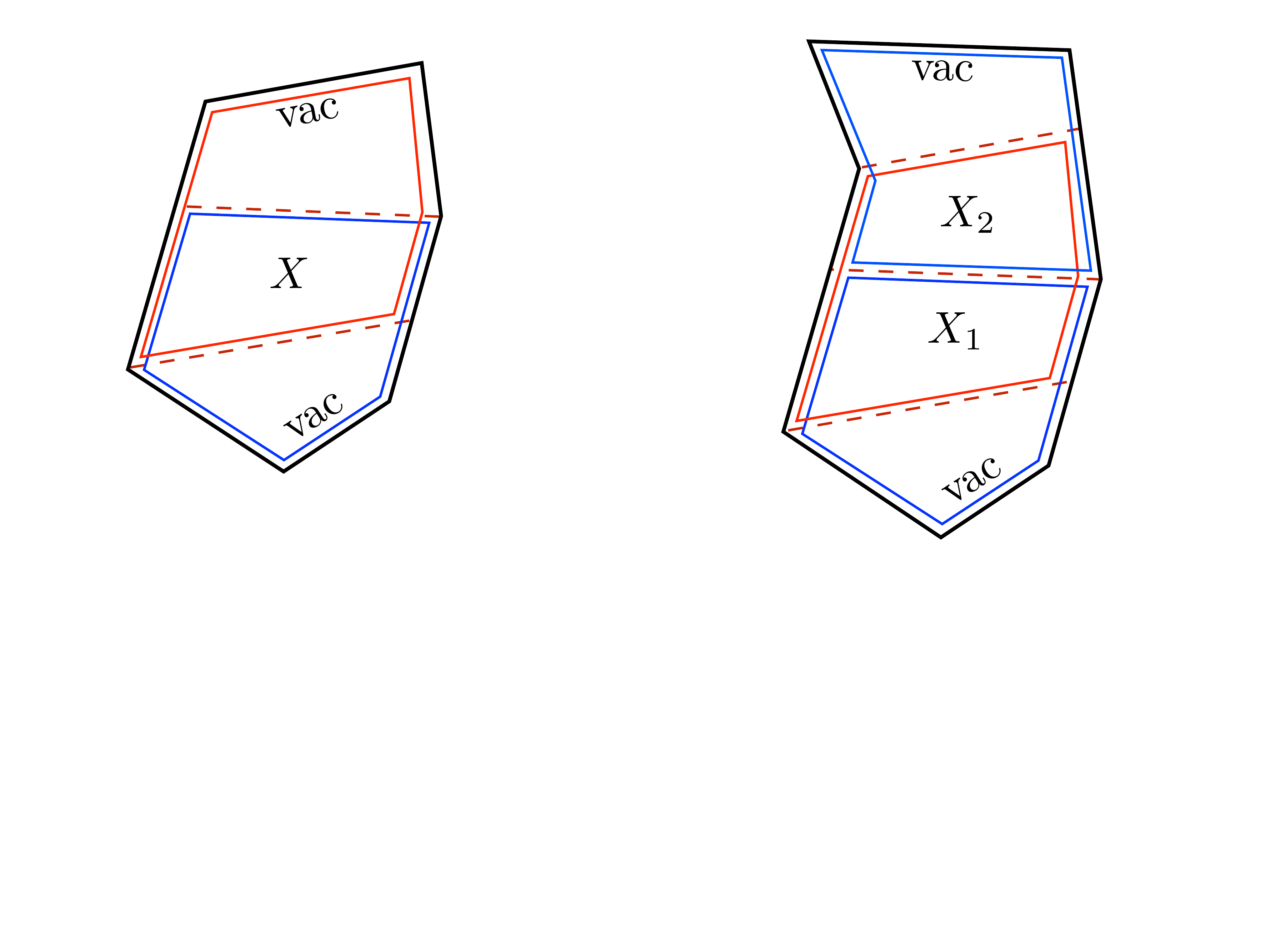}
\end{center}
\vspace{-4cm}
\caption{Decomposition of hexagon and heptagon Wilson loops into overlapping sequences of pentagons. For the hexagon, we have two pentagons overlapping on one square and 
correspondingly only one complete sum over intermediate states is needed. The heptagon has an extra hat, one more pentagon and middle square, and two sums are needed.}\label{hexhep}
\end{figure}

\subsection{Hexagon at weak coupling}

We begin with the hexagonal Wilson loop. It is convenient to use the 4d cross ratios $(u_{1}, u_{2}, u_{3})$ to parameterize its geometry. The latter can then be converted to the standard OPE 
parameters $(\tau, \sigma, \phi)$ through the map \cite{Gaiotto:2011dt,Basso:2013aha}
\beq\label{OPEpar}
u_{2} = \frac{e^{-2\tau}}{1+e^{-2\tau}}\, , \qquad u_{3} = \frac{1}{1+e^{2\sigma}+2\cos{\phi} \, e^{\sigma-\tau}+e^{-2\tau}}\, , \qquad u_{1} =  e^{2\sigma+2\tau} u_{2}u_{3}\, .
\eeq
The collinear limit corresponds to $\tau\rightarrow \infty$, at fixed flux tube position $\sigma$ and angle $\phi$; equivalently, $u_{2}\rightarrow 0$ with $u_{1}+u_{3} = 1$. The restriction to the 3d
kinematics is obtained by setting $\phi = 0$.%
\footnote{Another possible choice is $\phi = \pi$. We shall not consider it here.}

The OPE does not compute the vev of a Wilson loop, that is UV divergent, but instead a certain ratio $\mathcal{W}_{n}$ of Wilson loops, that is finite. The ratio is defined for given a tessellation of 
the loop in terms of pentagons, as shown in figure \ref{hexhep}. For instance, for the hexagon, it reads
\beq\label{OPEr}
\mathcal{W}_{6} = W_{6} \times \frac{W^{\textrm{m}}_{4}}{W^{\textrm{b}}_{5}\times W^{\textrm{t}}_{5}}\, ,
\eeq
where $W^{\textrm{b/t}}_{5}$ is the bottom/top pentagon WLs embedded in the hexagon and $W^{\textrm{m}}_{4}$ being the middle square Wilson loop on which the above two
pentagons overlap. This combination has the effect of subtracting the BDS component of the Wilson loop and replacing it by the abelian OPE ratio function \cite{Gaiotto:2011dt}. The 
latter is a finite function of the cross ratios (\ref{OPEpar}),
\beq
\begin{aligned}
\mathcal{W}_{6}^{\textrm{U}(1)} = \exp{\bigg[\frac{\Gamma_{\textrm{cusp}}}{4} r_{6}(\sigma, \tau, \phi)\bigg]} \, ,
\end{aligned}
\eeq
where
\beq\label{abelian}
\begin{aligned}
r_{6} = 2\zeta(2) - \log{(1-u_{2})}\log{\frac{u_{1}u_{2}}{u_{3}(1-u_{2})}}-\log{u_{1}}\log{u_{3}}-\sum_{i=1}^{3} \textrm{Li}_{2}(1-u_{i})\, ,
\end{aligned}
\eeq
and $\Gamma_{\textrm{cusp}}(g) = 2g^2 + O(g^4)$ is the cusp anomalous dimension. With its help, one can write
\beq\label{W6}
\mathcal{W}_{6} = \mathcal{W}_{6}^{\textrm{U}(1)}\times \mathcal{R}_{6}\, ,
\eeq
where $\mathcal{R}_{6} = 1+O(g^3)$ is the remainder function. So defined, the loop admits a nice expansion in the collinear limit, organized in terms of the twists of particles which 
are being exchanged between the bottom and top pentagons.

In the following, we will consider the leading twist-1 and twist-2 components only. They follow immediately from the large $\tau$ expansion of (\ref{abelian}), using the cross ratios 
(\ref{OPEpar}) and setting $\phi = 0$. The result reads
\beq\label{data6}
\begin{aligned}
\mathcal{W}_{6} = 1 &- g^2e^{-\tau} \left(e^{\sigma}\log{(1+e^{-2\sigma})}+e^{-\sigma}\log{(1+e^{2\sigma})} \right) \\
&+g^2 e^{-2\tau} \left(\sigma-\frac{1}{2}+\sinh{\sigma}(e^{\sigma}\log{(1+e^{-2\sigma})}-e^{-\sigma}\log{(1+e^{2\sigma})}) \right)\\
&+O(g^2 e^{-3\tau}, g^{3})\, .
\end{aligned}
\eeq 

\begin{figure}
\begin{center}
\includegraphics[scale=0.40]{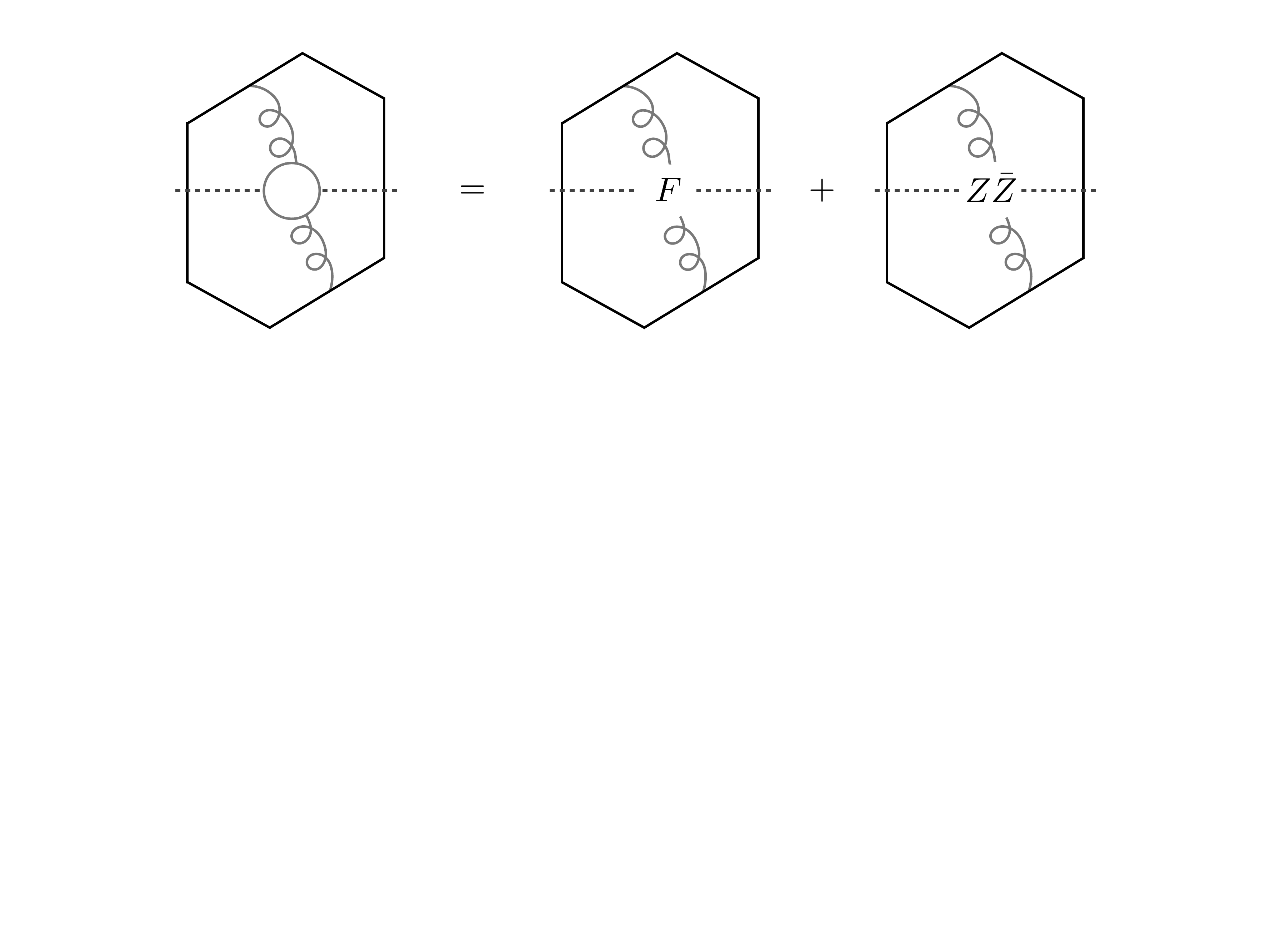}
\end{center}
\vspace{-7cm}
\caption{The OPE cut of the two loop contribution to the bosonic Wilson loop reveals a pair of matter fields, coming from the bubble correction to the Chern-Simons propagator. Its twist-1 
component can take the form of a gluonic flux tube excitation $F$ or of a spinon-anti-spinon pair $Z\bar{Z}$. They both start contributing at order $O(g^2)$ at weak coupling, according to 
our pentagon conjectures.}\label{OPEcut}
\end{figure}

One the flux tube side, the `1' in (\ref{data6}) comes from the vacuum state, while the next two terms from the twist-1 and twist-2 excitations, respectively. In the SYM theory, there is only one 
candidate at leading twist, the twist-1 gluon (which comes with two helicities). Everything else is either heavier or carries an R-charge. In the ABJM theory, we have a gluonic twist-1 excitation 
as well but we can also form a singlet combination of twist-1/2 spinons. We expect that both will contribute at two loops, since they should both stem from the collinear limit of a gluon propagator 
dressed  by a loop of hypermultiplets, as depicted in figure \ref{OPEcut}. This is what we are set to show below.

\subsubsection{Twist 1}

The gluon contribution to the hexagon is by far the simplest one to evaluate. It is given by the Fourier transform of the gluon measure,
\beq\label{WF}
\I_{F} = \int\frac{du}{2\pi} \mu_{F}(u) e^{-\tau E_{F}(u) + i\sigma p_{F}(u)}\, .
\eeq
To leading order in weak coupling, we have $E_{F}(u) = 1$, $p_{F}(u) = 2u$ and the measure takes the simple form (\ref{muFw}). We then evaluate the integral by closing the contour of integration in the upper half plane, summing up the residues. This yields 
\beq\label{WF1}
\I_{F} = -g^2 e^{-\tau} \times \frac{\sigma}{\sinh{\sigma}} + O(g^4)\, .
\eeq
This expression has poles at $e^{2\sigma} = 1$ in conflict with the expected analytical properties of the WL, see e.g.~the exact expression (\ref{data6}). Similar poles were uncovered in individual 
flux tube components of the SYM hexagonal WL at higher twists \cite{Hatsuda:2014oza,Drummond:2015jea}. They were observed to cancel out, however, after adding up all contributions at a 
given twist order. We expect a similar phenomenon to occur in the current situation. Namely, we regard the spurious poles in (\ref{WF1}) as an indication that other flux tube contributions must 
be added to the mix. The only candidate is a spinon--anti-spinon pair.

The $Z\bar{Z}$ contribution naturally take on the form of a two-fold integral over the two-spinon phase space,
\beq
\I_{Z\bar{Z}} = \int\frac{du_{1}du_{2}}{(2\pi)^2} \mu_{Z\bar{Z}}(u_{1}, u_{2}) e^{-\tau (E_{Z}(u_{1})+E_{\bar{Z}}(u_{2})) + i\sigma(p_{Z}(u_{1})+p_{\bar{Z}}(u_{2}))}\, .
\eeq
The energy and momentum are the same for the spinon and the anti-spinon and are given to leading order at weak coupling by Eq.\ (\ref{EZ1}). The rest of the integrand is assumed to take the 
factorized form introduced earlier and can be written as
\beq\label{mupair}
\mu_{Z\bar{Z}}(u_{1}, u_{2}) = \frac{-4\mu_{Z}(u_{1})\mu_{\bar{Z}}(u_{2})}{((u_{1}-u_{2})^2+4)P_{Z|\bar{Z}}(u_{1}|u_{2})P_{\bar{Z}|Z}(u_{2}|u_{1})}\, ,
\eeq
where $\mu_{Z} = \mu_{\bar{Z}}$ and $P_{Z|\bar{Z}} = P_{\bar{Z}|Z} = \bar{P}$ are the spinon measure and transitions considered in section \ref{Sub:spinons}. The rational part is needed to 
project the quantum numbers of the pair to the SU$(4)$ singlet channel. The overall minus was fixed a posteriori, such as to permit a successful comparison with the field theory answer. (The 
sign might look awkward but it would also be needed in the SYM theory if we were to consider the fermion--anti-fermion contribution to the hexagon using the normalization of Ref.\ 
\cite{Basso:2015rta} for the pentagons.)

Plugging the ans\"atze (\ref{Mansatz}) and (\ref{Wansatz}) for the weak coupling measure and transition into Eq, (\ref{mupair}), it yields
\beq\label{muZZb}
\mu_{Z\bar{Z}}(u_{1}, u_{2}) = \frac{-4\pi^2g^2 \cosh{\left(\tfrac{1}{2}\pi(u_{1}-u_{2}) \right)}}{((u_{1}-u_{2})^2+4)\cosh{(\pi u_{1})}\cosh{(\pi u_{2})}}+ o(g^2)\, .
\eeq
Notice that 1) the unpleasant square roots predicted by (\ref{Mansatz}) and (\ref{Wansatz}) combine together such that the resulting integrand is meromorphic and 2) the undermined factor $\z$ 
cancels out between the $\mu$'s in the numerator and the $\bar{P}$'s in the denominator. The integrand is of order $O(g^2)$ in agreement with the diagrammatic intuition, see figure \ref{OPEcut}.

We compute the integral by closing the contours in the upper half-planes and summing up the residues. Integrating first over $u_{2}$, we pick up the residues at $u_{2} = i/2+in$ and $u_{2} = u_{1}+2i$, with $n \in \mathbb{N}$, and then at $u_{1} = i/2+im$, with $m\in \mathbb{N}$. Thanks to 
the zeros in the numerator, only the residues corresponding to the odd powers of $e^{-\sigma}$ survive, in agreement with the structure of the perturbative answer (\ref{data6}). Combining 
everything together, we obtain
\beq
\I_{Z\bar{Z}} = -g^2 e^{-\tau}\times \left(2\cosh{\sigma} \log{(1+e^{-2\sigma})} - \frac{\sigma e^{-2\sigma}}{\sinh{\sigma}} \right)\, .
\eeq
It displays the same spurious poles as the gluon part. They readily cancel up in the sum, as anticipated,
\beq
\I_{F} + \I_{Z\bar{Z}} = -g^2 e^{-\tau} \left(e^{\sigma}\log{(1+e^{-2\sigma})}+e^{-\sigma}\log{(1+e^{2\sigma})} \right) + o(g^2)\, .
\eeq
This is precisely the field theory result (\ref{data6}). Interestingly, although the OPE representation discussed here is more involved compared to the one in SYM, --- we have a double integral 
at leading twist in the ABJM case, --- the final expression ends up being the same as in the SYM theory at one loop, up to a factor $1/2$ to accommodate the difference in the cusp anomalous 
dimensions in the two theories. We also note that the bulk of the final answer comes from the $Z\bar{Z}$ pair.

\subsubsection{Twist 2}

There are many more states to consider at twist-2 level. The complete list includes
\beq
DF, \quad FF, \quad FZ\bar{Z}, \quad \Psi\Psi, \quad Z^2\Psi, \quad \bar{Z}^2\Psi, \quad Z^4, \quad \bar{Z}^4, \quad (Z\bar{Z})^2\, ,
\eeq
where $DF = F_{2}$ is the twist-2 gluon bound state, $FF$ a two-gluon state, etc. However, if our ans\"atze are correct, assuming also that $\z = O(g^0)$, then only 4 of the above states 
contribute at order $O(g^2)$, namely, $DF, \Psi\Psi, \Psi Z^2$ and $\Psi \bar{Z}^2$.%
\footnote{We find that $FF, FZ\bar{Z}, (Z\bar{Z})^2, Z^4$ scale as $g^8, g^8, g^8/\alpha^{8},g^{8}/\alpha^{16}$, respectively.}

The gluon contribution is again the easiest one to write. It follows directly from (\ref{muFa}),
\beq
\I_{DF} = g^2 e^{-2\tau} \int \frac{du}{2\pi} e^{2iu \sigma} \frac{\pi^2 u^2}{2\sinh^2{(\pi u)}} 
= 
g^2 e^{-2\tau}\frac{\sigma (e^{\sigma}+e^{-\sigma})-(e^{\sigma}-e^{-\sigma})}{(e^{\sigma}-e^{-\sigma})^3} \, .
\eeq
As before, it has again the undesired singularities at $e^{2\sigma} = 1$.

Then comes the $\Psi\Psi$ contribution
\beq\label{psipsi}
\I_{\Psi\Psi} = \frac{1}{2}e^{-2\tau}\int \frac{du_{1}du_{2}}{(2\pi)^2} \frac{6\mu_{\Psi}(u_{1})\mu_{\Psi}(u_{2}) e^{i(p_{\Psi}(u_{1})+p_{\Psi}(u_{2}))\sigma}}{((u_{1}-u_{2})^2+4)((u_{1}-u_{2})^2+1)P_{\Psi|\Psi}(u_{1}|u_{2})P_{\Psi|\Psi}(u_{2}|u_{1})}\, ,
\eeq
with a symmetry factor in front compensating for the two identical fermions. The matrix part is as for the two-scalar contribution to the SYM hexagon \cite{Basso:2014koa}. Looking at the weak 
coupling formulae 
(\ref{PPsi}) and (\ref{muPsi}) for the fermion pentagon and measure, one would conclude that this integral is $\sim g^8$ at weak coupling, that is 8 loops in the ABJM theory. This estimate is not  
correct however. It overlooks the fact that the fermions develop quite a strange behavior at small momenta, i.e., $p\sim g^2$, and the aforementioned weak coupling formulae do not properly represent
this domain. What we need instead are the weak coupling expressions on the so-called small fermion sheet. They are obtained through an analytic continuation using formulae at finite coupling, 
as described in Ref.\ \cite{Basso:2014koa}. The small fermion sheet, reached via the above procedure, can be parameterized in terms of a rapidity $u$, with $u = \infty$ corresponding to zero 
momentum. Following \cite{Basso:2014koa}, we will denote functions evaluated on that sheet, like the momentum, the energy, etc., with a `check' on the rapidity, e.g.,
\beq\label{small-p}
p_{\Psi}(\check{u}) = 2g^2/u +O(g^4)\, , \qquad E_{\Psi}(\check{u}) = 1+O(g^6)\, .
\eeq
Other quantities like the measure and pentagon transitions also drastically simplify. In particular, one finds, after folding the 4d formulae in appendices of Ref.\ \cite{Basso:2015rta} into 3d ones,
\beq\label{Psmall}
\frac{1}{P_{\Psi|\Psi}(\check{u}_{2}|u_{1})P_{\Psi|\Psi}(u_{1}|\check{u}_{2})} = u_{2}^2 + O(g^2)\, ,
\eeq
together with
\beq\label{small-mu}
\mu_{\Psi}(\check{u}_{2}) = -1 + O(g^2)\, .
\eeq
We cannot have more than one small fermion at a time in the case at hand, since to produce a non-vanishing contribution, the small fermions must always bind to something `big'. Here, 
one fermion 
will attach to the other and form a string; of course, it does not matter which one we choose, as long as we add a factor $2$ in the end to reflect the doubling. So we can use (\ref{Psmall}), 
(\ref{small-mu}) in equation (\ref{psipsi}) as well as (\ref{muPsi}) for the measure of the large fermion $\Psi(u_{1})$. The resulting integrand is then of order $O(g^2)$ as desired.

We can then integrate the small fermion out by attaching it to the other one. The string is determined by the zeros of the rational factor in (\ref{psipsi}). Here we get two options, 
$u_{2} = u_{1} - i$ and $u_{2} = u_{1}-2i$.%
\footnote{The contour of integration in the small fermion domain goes anti-clockwise around all singularities in the lower half plane, see \cite{Basso:2015rta} for further detail.}
Picking up the residues, we arrive at
\beq
\begin{aligned}
\I_{\Psi\Psi} = \frac{g^2e^{-2\tau}}{2}\int_{\mathbb{R}+i0} \frac{du_{1}}{2\pi} \frac{\pi^2(u_{1}^2+2)}{\sinh^2{(\pi u_{1})}} e^{2iu_{1}\sigma} 
= 
g^2 e^{-2\tau}\bigg[ \frac{\sigma(2-5e^{2\sigma}+e^{4\sigma})}{(1-e^{2\sigma})^3}-\frac{e^{2\sigma}}{(1-e^{2\sigma})^2}\bigg] \, .
\end{aligned}
\eeq
The $i0$ prescription is a remnant of the splitting of the kinematics into the small and large domains, see \cite{Basso:2014koa}, and is needed to avoid the double pole at $u_{1}=0$. (More 
precisely, the latter pole is a trace of the small-fermion region that collapses into a point on the large fermion sheet.) We notice here again the presence of unwanted singularities. 

Finally, we have the integral for $\Psi ZZ$, and equivalently $\Psi \bar{Z}\bar{Z}$,
\beq\label{PsiZZ}
\I_{\Psi ZZ} = \frac{1}{2}e^{-2\tau}\int \frac{du dv_{1} dv_{2}}{(2\pi)^3} 
\frac{e^{i(p_{\Psi}(u)+p_{Z}(v_{1})+p_{Z}(v_{2}))\sigma}\mu_{\Psi}(u)\mu_{Z}(v_{1})\mu_{Z}(v_{2}) \Pi_{\Psi ZZ}}{P_{Z|Z}(v_{1}|v_{2})P_{Z|Z}(v_{2}|v_{1})
\prod_{i=1,2}P_{\Psi|Z}(u|v_{i})P_{Z|\Psi}(v_{i}|u)}\, ,
\eeq
with an overall symmetry factor removing overcounting due to the identity of the spinons. The matrix part $\Pi_{\Psi ZZ}$ can be obtained from the integral formula of \cite{Basso:2015uxa} 
or by contracting the matrix pentagons of \cite{Belitsky:2016vyq}. For the singlet channel in $\textbf{6}\otimes \textbf{4}\otimes \textbf{4}$, it yields  
\beq\label{mPsiZZ}
M_{\Psi(u) Z(v_{1})Z(v_{2})} =  \frac{12}{\left((u-v_{1})^2+\frac{9}{4}\right)\left((u-v_{2})^2+\frac{9}{4}\right)((v_{1}-v_{2})^2+1)}\, .
\eeq
The spinon part of the integrand reads, according to (\ref{Wansatz}) and 
(\ref{Mansatz}),
\beq\label{intZZ2}
\frac{\mu_{Z}(v_{1})\mu_{Z}(v_{2})}{P_{Z|Z}(v_{1}|v_{2})P_{Z|Z}(v_{1}|v_{2})} = \frac{\pi^2 g^2 (v_{1}-v_{2})
\sinh{\left(\tfrac{1}{2}\pi(v_{1}-v_{2})\right)}}{\z^4 \cosh{(\pi v_{1})}\cosh{(\pi v_{2})}}+ o(g^2)\, .
\eeq
Hence, after using the fermion data (\ref{muPsi}) and (\ref{PFZ}), the integrand is superficially small, of order $O(g^{8}/\alpha^4)$. However, here again the dominant contribution does not 
come from the kinematical domain where the latter formulae apply, but from the small fermion domain. Continuing our expressions to that sheet and taking the weak coupling limit afterwards, 
one obtains
\beq\label{inZPsi}
\frac{1}{P_{\Psi|Z}(\check{u}|v)P_{Z|\Psi}(v|\check{u})} = u + O(g^2)\, .
\eeq
Together with (\ref{small-mu}) it takes out six powers of $g$ and returns an integrand of order $O(g^2/\alpha^4)$. The poles in the matrix part (\ref{mPsiZZ}) dictate that the small fermion 
binds below the spinon's rapidity $v_i$ at $u = v_{1,2}-3i/2$. Picking these residues up and using (\ref{intZZ2}) for the rest, it yields
\beq
\I_{\Psi ZZ} = g^2e^{-2\tau}\int \frac{dv_{1} dv_{2}}{(2\pi)^2} \frac{\pi^2(9+2v^2_{1}+2v_{2}^2) (v_{1}-v_{2})\sinh{\left(\tfrac{1}{2}\pi(v_{1}-v_{2})\right)}
e^{i(v_{1}+v_{2})\sigma}}{\z^4((v_{1}-v_{2})^2+1)((v_{1}-v_{2})^2+9)}\,.
\eeq
We then simply repeat the analysis carried out earlier for the two spinon integral and find
\beq
\begin{aligned}
&\I_{\Psi ZZ}  =\\
& \frac{g^2 e^{-2\tau}}{2\z^4} 
\bigg[\frac{-1+6e^{2\sigma}-e^{4\sigma}}{2(1-e^{2\sigma})^2}+\frac{\sigma e^{2\sigma}(-1+9e^{2\sigma}-5e^{4\sigma}+e^{6\sigma})
}{
(1-e^{2\sigma})^3}+\frac{1}{2}(e^{\sigma}-e^{-\sigma})^2\log{(1+e^{2\sigma})}\bigg] \, .
\end{aligned}
\eeq

Adding everything up, one verifies that the bad singularities go away and that the sum matches with (\ref{data6}) if $\z^4 = 1$.\\

\subsubsection{Four loop leading discontinuity}

Being convinced that our formulae work correctly at weak coupling, at least at low twists, we can use them to make higher loop predictions for the leading OPE discontinuities (LD) 
\cite{Alday:2010ku,Gaiotto:2010fk}. The latter correspond to terms exhibiting maximal powers of the OPE time $\tau$ at a given loop order. They follow unambiguously from dressing 
the flux tube integrands with the leading weak coupling corrections to the energies of the flux-tube excitations. Realizing that these corrections all start at two loops, we obtain
\beq
\mathcal{W}_{6}(\sigma, \tau)\big|_{\textrm{LD}} = \sum_{L=2}^{\infty} g^{2L} \tau^{L-1} (e^{-\tau}f^{(1)}_{L}(\sigma) + e^{-2\tau}f^{(2)}_{L}(\sigma) + \ldots )\, ,
\eeq
where $f^{(n)}_{L}(\sigma)$ is a coupling independent function of $\sigma$. We focus here on the LD $\propto g^{4}\tau$.

At leading twist, plugging into (\ref{WF}) the correction (\ref{EF1}) to the energy of a gluon, and expanding the exponent at weak coupling, provides the gluon contribution to $f^{(1)}_{2}$,
\beq
\begin{aligned}
f_{2}^{(1)}|_{F} &= \int\frac{\pi du}{\cosh^2{(\pi u)}} e^{2iu \sigma} \left(\psi(\tfrac{3}{2}+iu)+\psi(\tfrac{3}{2}-iu)-2\psi(1)\right) \,.
\end{aligned}
\eeq 
Similarly, one gets with (\ref{EZ1}) the $Z\bar{Z}$ contribution,
\beq
\begin{aligned}
f_{2}^{(1)}|_{Z\bar{Z}} &= 2\int du_{1}du_{2} 
\frac{\cosh{\left(\tfrac{1}{2}\pi(u_{1}-u_{2})\right)}\left(\psi(\tfrac{1}{2}+iu_{1})+\psi(\tfrac{1}{2}-iu_{1})-2\psi(1)\right)}{((u_{1}-u_{2})^2+4)\cosh{(\pi u_{1})}\cosh{(\pi u_{2})}} e^{i(u_{1}+u_{2})\sigma}\, .
\end{aligned}
\eeq
The integrals can be evaluated by picking up the residues. Then their sum can be expressed in a concise form as
\beq
f_{2}^{(1)} = e^{\sigma} \log{(1+e^{-2\sigma})}(2-\log{(1+e^{2\sigma})})+e^{-\sigma} \log{(1+e^{2\sigma})}(2-\log{(1+e^{-2\sigma})})\, .
\eeq
Remarkably, it is identical to the LD of the two loop hexagon (at $\phi = 0$) in SYM up to a factor of $1/4$.

We proceeded similarly at twist 2 and found that here again the result coincides with the SYM expression. Extrapolating to higher twist, one can reasonably conjecture that the LD of the four 
loop hexagonal WL in the ABJM is $1/4$ the corresponding two-loop LD in the SYM theory. (Explicit expression for the LD of the SYM hexagon can be extracted from the formulae given 
in \cite{Gaiotto:2010fk}.) It would be interesting to further test this extrapolation and see if one can `bootstrap' the missing 4 loop information -- in the spirit of what was done in 
\cite{Gaiotto:2010fk} for the 2 loop hexagon WL in the SYM theory or at higher loops using the hexagon function bootstrap program \cite{Dixon:2011pw,Dixon:2013eka,Dixon:2014voa,Caron-Huot:2016owq}. 

\subsection{Heptagon at weak coupling}

We can probe more of the pentagon transitions by considering the heptagon WL, shown in the right panel of figure \ref{hexhep}. After modding out by the pentagons and squares in the 
sequence, the OPE ratio reads
\beq
\mathcal{W}_{7} = \exp{\left(r_{6}(\sigma_{1}, \tau_{1}, \phi_{1})+ r_{6}(\sigma_{2}, \tau_{2}, \phi_{2}) + r_{7}(\sigma_{1}, \sigma_{2}, \tau_{1}, \tau_{2}, \phi_{1}, \phi_{2}) \right)}\times \mathcal{R}_{7}\, ,
\eeq
with the restriction to the 3d kinematics corresponding to $\phi_{1,2} = 0$. According to \cite{Wiegandt:2011zz}, the  remainder function $\mathcal{R}_{7}  =1+ O(g^3)$, and thus $\mathcal{W}_7$ 
should match with the SYM answer to leading order at weak coupling $\sim g^2$. The $r_{6}$ components originate from hexagons embedded inside the heptagon and their OPE analysis reduces 
to the one carried out earlier. The interesting new ingredient is the 7-point abelian remainder function $r_{7}$ that was constructed in \cite{Sever:2011pc}. It describes flux-tube excitations traveling 
all the way from the bottom to the top of the heptagon; $X_{1}, X_{2} \neq \emptyset$ in figure \ref{hexhep}. It is a function of two OPE time $\tau_{1,2}$ and space $\sigma_{1,2}$ coordinates. We 
shall only consider it to leading order in the double collinear limit $\tau_{1, 2}\rightarrow \infty$ which flattens the heptagon on the middle pentagon in figure \ref{hexhep}. The relevant expression is
\beq\label{r7LO}
\begin{aligned}
\mathcal{W}_{7}|_{\textrm{conn}} = 1+ g^2 e^{-\tau_{1}-\tau_{2}}\bigg[&e^{\sigma_{1}+\sigma_{2}}\log{\frac{(1+e^{2\sigma_{1}})(1+e^{2\sigma_{2}})}{e^{2\sigma_{1}}
+
e^{2\sigma_{2}}+e^{2\sigma_{1}+2\sigma_{2}}}} \\
&+ 2e^{\sigma_{1}-\sigma_{2}}\log{\frac{e^{2\sigma_{1}}(1+e^{2\sigma_{2}})}{e^{2\sigma_{1}}+e^{2\sigma_{2}}+e^{2\sigma_{1}+2\sigma_{2}}}} + \sigma_{1}\leftrightarrow \sigma_{2} \bigg]+ \ldots\, ,
\end{aligned}
\eeq
where the ellipses stand for higher twist corrections. It is obtained from the expression analyzed in \cite{Sever:2011pc,Basso:2013vsa,Basso:2013aha} by setting $\phi_{1} = \phi_{2} = 0$.

On the flux tube side, there are 4 distinct processes contributing to (\ref{r7LO}) at leading twist in the bottom and top channels, namely
\beq
\begin{aligned}
&1) \qquad  \textrm{vacuum} \rightarrow F(u) \rightarrow F(v) \rightarrow \textrm{vacuum}\, , \\
&2)\qquad \textrm{vacuum} \rightarrow F(u) \rightarrow Z(v_{1})\bar{Z}(v_{2}) \rightarrow \textrm{vacuum}\, , \\
&3)\qquad \textrm{vacuum} \rightarrow Z(u_{1})\bar{Z}(u_{2}) \rightarrow F(v) \rightarrow \textrm{vacuum}\, , \\
&4)\qquad \textrm{vacuum} \rightarrow Z(u_{1})\bar{Z}(u_{2}) \rightarrow Z(v_{1})\bar{Z}(v_{2}) \rightarrow \textrm{vacuum}\, .
\end{aligned}
\eeq
Process 1) parallels the one studied in \cite{Basso:2013vsa,Basso:2013aha} for the SYM theory. The integrand is given by
\beq\label{I1}
\I_{F|F} = e^{-\tau_{1}-\tau_{2}}\int \frac{du dv}{(2\pi)^2} \mu_{F}(u)\mu_{F}(v) P_{F|F}(-u|v) e^{ip_{F}(u)\sigma_{1}+ip_{F}(v)\sigma_{2}}\, ,
\eeq
where the contour of integration is taken to be $\mathbb{R}+i0$ in both cases. The prescription is needed to avoid the decoupling pole at $u = -v$ and is dictated by the kinematics of 
the heptagon WL, see discussion in \cite{Basso:2013aha}.%
\footnote{The contours are such that the heptagon integral (\ref{I1}) reduces to the hexagon one (\ref{WF}) when $\sigma_{1,2}\rightarrow -\infty$ and $\sigma = \sigma_{2}-\sigma_{1}$ 
is held fixed.} At weak coupling, we replace the momenta by twice their arguments and use expressions (\ref{PFw}) and (\ref{muFw}) for the pentagon and measure,
\beq
 \mu_{F}(u) P_{F|F}(-u|v)\mu_{F}(v) 
 = 
 \frac{-g^2\Gamma^2\left(\tfrac{3}{2}+iu\right)\Gamma(-iu-iv)\Gamma(2-iu-iv)\Gamma^2\left(\tfrac{3}{2}+iv\right)}{\left(u^2+\tfrac{1}{4}\right)\left(v^2+\tfrac{1}{4}\right)}\, .
\eeq
The integrand is of order $O(g^2)$ as expected. Evaluating the integral by picking up residues in the upper half planes, we obtain
\beq
\I_{F|F} = -2g^2 e^{-\sum_i(\tau_{i}+\sigma_{i})}\big[ 1 +(2-3\sigma_{1})e^{-2\sigma_{1}} +(2-3\sigma_{2})e^{-2\sigma_{2}} + \ldots \big]\, ,
\eeq
for the first few terms in the asymptotic limit $\sigma_{1,2} = \infty$. Processes 2) and 3) are symmetrical and can be obtained from one another by permuting the OPE coordinates 
$\sigma_{1,2}\rightarrow \sigma_{2,1}$. The integral for 2) is given by
\beq
\I_{F|Z\bar{Z}} = e^{-\tau_{1}-\tau_{2}}\int \frac{du dv_{1}dv_{2}}{(2\pi)^3} e^{ip_{F}(u)\sigma_{1}+i\sum_i p_{Z}(v_{i})\sigma_{2}}\mu_{F}(u)\mu_{Z\bar{Z}}(v_{1}, v_{2})\prod_{i}P_{F|Z}(-u|v_{i})\, ,
\eeq
with the two-spinon measure~(\ref{muZZb}). There is no decoupling pole to handle here and thus the integrals can be taken directly along the real lines. Using (\ref{PFZ}) for the 
gluon-to-spinon pentagon, one verifies that the integrand is of order $O(g^2)$ and one easily obtains
\beq
\mathcal{W}_{F|Z\bar{Z}} = g^2 e^{-\sum_{i}(\sigma_{i}+\tau_{i})}\left[ 1 +(3-4\sigma_{1})e^{-2\sigma_{1}} +\frac{3}{2}(3-4\sigma_{2})e^{-2\sigma_{2}} + \ldots \right]\,.
\eeq
\begin{figure}
\begin{center}
\includegraphics[scale=0.45]{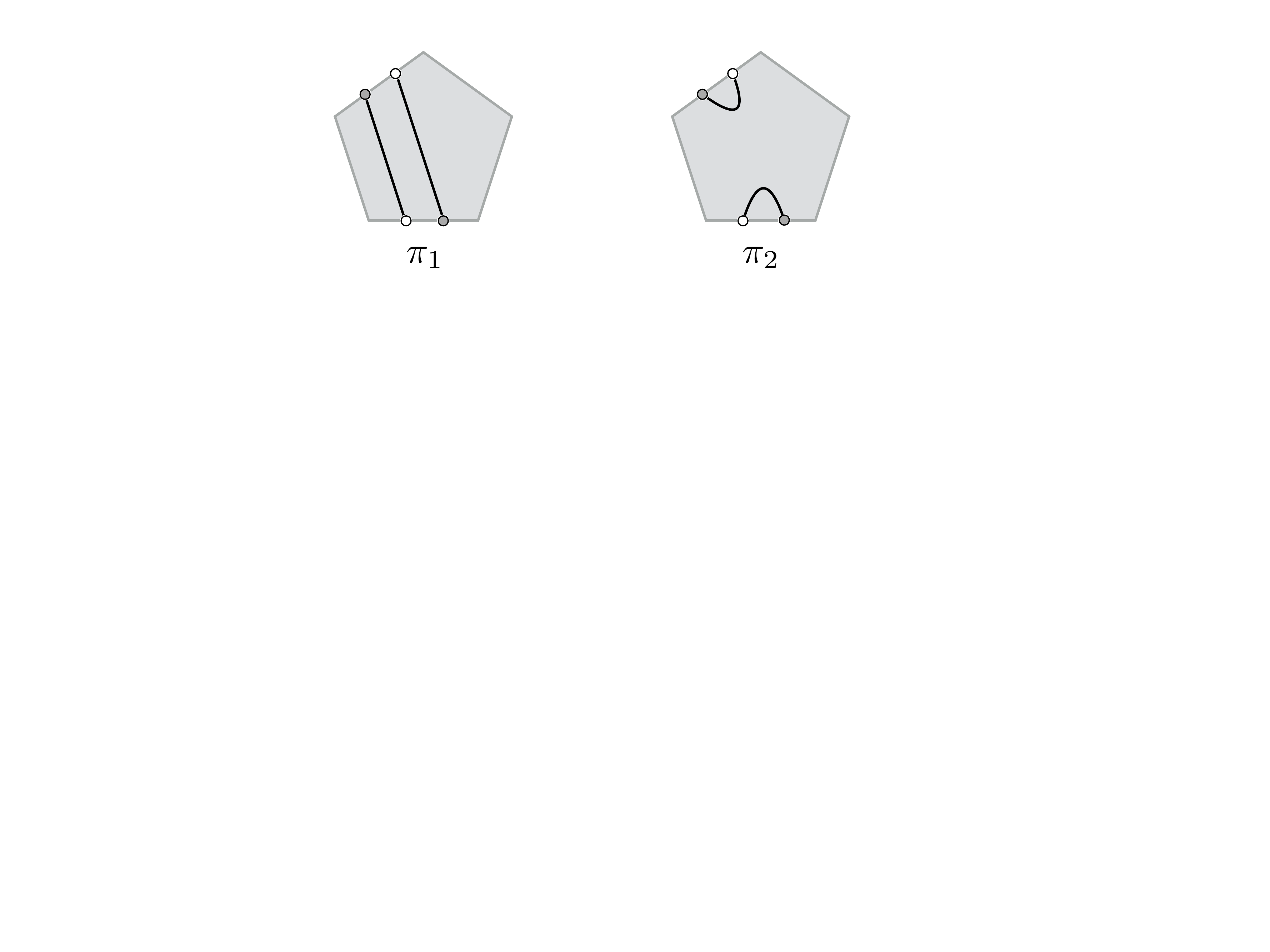}
\end{center}
\vspace{-9cm}
\caption{Cartoon of the matrix structure of the $Z_{A}\bar{Z}^{B}\rightarrow Z_{C}\bar{Z}^{D}$ transitions. There are two structures for the two possible ways of contracting 
indices, $\pi_{1}\delta_{A}^{C}\delta_{D}^{B}+\pi_{2} \delta_{A}^{B}\delta_{D}^{C}$.}\label{mp}
\end{figure}
The final process involves a non-trivial transition between two $Z\bar{Z}$ pairs at the bottom and top of the pentagon. The integrand is given by
\beq
\mu_{Z\bar{Z}}(u_{1}, u_{2})\mu_{Z\bar{Z}}(v_{1}, v_{2})\times \prod_{i}P_{Z|Z}(-u_{i}|v_{i})P_{Z|\bar{Z}}(-u_{i}|v'_{i}) \times M(\{-u\}, \{v\}) \, ,
\eeq
where $v'_{1,2} = v_{2,1}$. It involves a nontrivial matrix part $M(\{-u\}, \{v\})$ which receives contributions from the two tensors allowed for the transition, see figure \ref{mp}. The two 
associated polynomials in rapidities differences can be found in \cite{Belitsky:2016vyq}. Here we need their sum in the singlet channel,
\beq
M(\{u\}, \{v\}) = (u_{1}-v_{1})(u_{2}-v_{2}+i)-\frac{1}{4}(u_{1}-u_{2}-2i)(v_{1}-v_{2}+2i)\, .
\eeq
Plugging this expression in the integrand and using our guesses for the spinon transitions (\ref{Wansatz}), we find that the integrand is meromorphic, of order $O(g^2)$, and that it does 
not depend on $\alpha$. We get
\beq
\mathcal{W}_{Z\bar{Z}|Z\bar{Z}} = -g^2 e^{-\sum_i(\tau_{i}+\sigma_{i})}\big[ 1 +(3-4\sigma_{1})e^{-2\sigma_{1}} +(3-4\sigma_{2})e^{-2\sigma_{2}} + \ldots \big]\, ,
\eeq
where integration is performed using $+i0$ prescriptions for the decoupling poles.

One can finally take the sum of all these terms and verify the agreement with the field theory result (\ref{r7LO}). We checked it up to high order in the double expansion at large $\sigma_{i}$'s.

\subsection{Strong coupling}

Complementary tests of our ans\"atze can be carried out at strong coupling. Wilson loops can be computed at strong using AdS minimal surfaces,
\beq
\log{\mathcal{W}} \simeq \mathcal{A} \, .
\eeq
Integrability greatly helps finding the minimal area $ \mathcal{A}$ for null polygonal contours and allows one to cast the answer in the form of the free energy of a system of Thermodynamic 
Bethe Ansatz (TBA) equations \cite{Alday:2009dv,Alday:2010vh}, see also \cite{Ito:2018puu,Hatsuda:2010cc} for recent studies. One can develop their systematic expansion in the near collinear regime \cite{Alday:2010ku}, which corresponds to the low 
temperature expansion of the TBA equations. Here we will only discuss the leading contributions for the hexagonal and heptagonal Wilson loops. They are controlled by the spectrum of AdS 
excitations, the TBA weights and TBA kernels. The expressions for AdS$_{4}$ can be straightforwardly obtained by folding those of the AdS$_{5}$ case.

Let us illustrate this for the hexagonal loop. In AdS$_{5}$, the renormalized minimal area receives two types of contributions at strong coupling, see \cite{Alday:2010vh,Alday:2010ku}, 
\beq
\mathcal{A}_{{\rm AdS}_{5}} = \Gamma^{\mathcal{N}=4}_{\textrm{cusp}}(g) \times 
\left( e^{i\phi}A_{\sqrt{2}}(\sigma, \tau) + A_{2}(\sigma, \tau)+e^{-i\phi}A_{\sqrt{2}}(\sigma, \tau) \right) + \ldots\, ,
\eeq
from two transverse modes with mass $\sqrt{2}$ and from one longitudinal mass 2 boson. The dots above stand for contributions of multi-particle states that will not be needed. The reduction 
to AdS$_{4}$ follows simply by setting $\phi = 0$ and adjusting the string tension,
\beq\label{AdS4}
\mathcal{A}_{{\rm AdS}_{4}} = \Gamma^{\mathcal{N}=6}_{\textrm{cusp}}(g)\times \left(2A_{\sqrt{2}}(\sigma, \tau) + A_{2}(\sigma, \tau) \right) + \ldots\, .
\eeq
Since at a given $g$, the cusp anomalous dimension in the ABJM theory is half the one of SYM, 
\beq\label{cusp-strong}
\Gamma^{\mathcal{N}=6}_{\textrm{cusp}}(g) = \frac{1}{2} \Gamma^{\mathcal{N}=4}_{\textrm{cusp}}(g) = g + O(1)\, ,
\eeq
we conclude that the contribution per unit of $g$ from a transverse mode is the same in the two theories. It implies on the flux tube side that the gluon measure $\mu_{F}$ should be identical 
to its SYM counterpart at strong coupling,
\beq
\mu_{F}|_{\mathcal{N}=6} = \mu_{F}|_{\mathcal{N}=4} + O(1/g)\, .
\eeq
This stringy prediction is easily seen to be obeyed by our formula (\ref{measure-gluon}) for the gluon measure, after using that $\bar{P}_{\mathcal{N}=4}(u|u) =1 +O(1/g)$, see, 
e.g., \cite{Basso:2014koa}.

The analysis for the mass 2 boson is more delicate. Like in the SYM theory \cite{Zarembo:2011ag,Basso:2014koa,Bonini:2015lfr,Bonini:2018mkg}, this boson does not correspond to a 
fundamental flux tube excitation at finite coupling. It is closer to a virtual bound state that reaches the two fermion threshold at strong coupling. As such it originates from the two fermion 
integral (\ref{psipsi}). In Appendix \ref{App:fermions} we show that the latter integral is half the corresponding one in the SYM theory at strong coupling in perfect agreement with the minimal surface prediction.

Finally we can verify our pentagon transition for the gluons by considering the heptagonal Wilson loop. The pentagon encodes information about the TBA kernel $K$ connecting neighbouring channels. The map is given by \cite{Basso:2013vsa}
\beq\label{PtoK}
P\ = 1 + \frac{1}{2g} K + \ldots\, ,
\eeq
and as written it can be applied to both the SYM and ABJM theory. On the TBA side, because of the folding relation, the kernel connecting transverse bosons is simply obtained by averaging over the two transverse modes,
\beq
K_{{\rm AdS}_{4}} = K_{{\rm AdS}_{5}} + \bar{K}_{{\rm AdS}_{5}}\, .
\eeq
Taking (\ref{PtoK}) into account, it agrees with the doubling relation (\ref{PN6}). Clearly, the doubling relation (\ref{PN6}) is the most natural all order uplift of this kernel averaging procedure that 
relates AdS$_{4}$ to AdS$_{5}$. This observation concludes the tests of our formulae on the Wilson loop side.

\section{Amplitudes}\label{Sec4}

While the application of the pentagon paradigm to the Wilson loop expectation values, described in the previous section, should not be surprising at all, in this section, we 
will extend it to the ABJM amplitudes. As we already alluded to in the introduction, the four-leg amplitude at lowest orders of perturbative series is identical to the 
four-cusp bosonic Wilson loop, hinting at an MHV-like duality previously unveiled in the SYM case. However, for the case at hand, it stops right there and 
begs for a supersymmetric extension to account for non-MHV amplitudes.

Since the $\mathcal{N} = 6$ supersymmetry is not maximal, the on-shell particle multiplet is not CPT self-conjugate and is packaged in two $\mathcal{N} = 3$ 
superfields,
\begin{align}
\Phi &
= \phi^4 + \theta^a \psi_a + \frac{1}{2} \epsilon_{abc} \theta^a \theta^b \phi^c + \frac{1}{3!} \epsilon_{abc} \theta^a \theta^b \theta^c \psi_4
\, , \\
\bar\Psi &
= \bar\psi^4 + \theta^a \bar{\phi}_a +\frac{1}{2} \epsilon_{abc} \theta^a \theta^b \bar\psi^c + \frac{1}{3!} \epsilon_{abc} \theta^a \theta^b \theta^c \bar{\phi}^4
\, , 
\end{align}
given by terminating expansions in the Grassmann variables $\theta^a$ with $a = 1,2,3$ and where $\epsilon_{abc}$ is the associated totally antisymmetric tensor. In this superspace 
representation, the SU(4) symmetry of the Lagrangian is broken down: the original R-symmetry index is split up as $A= (a,4)$ and only the U$(3)$ remains explicit. (Also, since the gauge 
fields are pure gauges, they do not emerge as asymptotic on-shell states.) Factoring out a super-delta function for super-momentum conservation and a Parke-Taylor-like prefactor  
\cite{Huang:2010qy}, the $n$-leg super-amplitude reads
\begin{align}
\label{SuperAmplitudeAn}
A_{n}(\bar\Psi_1 \Phi_2 \bar\Psi_3 \Phi_4 \dots \bar\Psi_{n-1} \Phi_n)
= \frac{\delta^{3} (P)\delta^{6}(Q)}{\sqrt{-\vev{12} \vev{23} \dots \vev{n1}}} \times \mathbb{A}_n(\theta)
\, ,
\end{align}
where $\mathbb{A}_n(\theta)$ is an observable that it similar in spirit to the super-loop in the SYM theory.

Several comments are in order with regards to this expression. First, the amplitude can have only an even number of external legs \cite{Agarwal:2008pu} as a consequence of alternating 
the gauge groups of the elementary fields along the color-ordered trace. Second, the $n$-point amplitude $A_n$ has Grassmann degree $\frac{3}{2}n$ and thus the reduced amplitude 
$\mathbb{A}_n$ inherits the residual degree $\frac{3}{2}(n-4)/2$ in $\theta$'s; it is N$^{\frac{1}{2}(n-4)}$MHV in four-dimensional terminology. Finally, dividing out the bosonic loop 
$W_{n}$ from the ``super-loop'' $\mathbb{A}_{n}$ should remove the divergences and return a dual-conformal invariant ratio
\begin{align} 
\label{RatioFunctionPn}
\mathbb{R}_n
=
\mathbb{A}_n / W_{n}\, .
\end{align}
However, despite its nice properties, this is not the object that naturally arises in the OPE. Instead, the OPE ratio is canonically defined by dividing by pentagons and multiplying by squares, as 
illustrated earlier, see equation (\ref{OPEr}). The ``super-loop'' $\mathbb{W}_n$, which we shall be analyzing below, is of this type. It can be built from $\mathbb{R}_n$ and the bosonic 
OPE ratio function $\mathcal{W}_{n}$,
\begin{align}
\label{WLobservable}
\mathbb{W}_n = \mathbb{R}_n \times \mathcal{W}_n\, .
\end{align}
The hexagon and heptagon $\mathcal{W}$ were discussed in the previous sections, however, only even $n$'s play a role in the consideration that follows. Notice that to leading order at weak coupling $\mathcal{W}_{n} = W_{n} = 1+O(g^2)$ and thus all these super-objects are identical at tree level and one loop, $\mathbb{W}_{n} = \mathbb{A}_{n} = \mathbb{R}_{n}$ when $g\rightarrow 0$.

Contrary to the Wilson loop expectation values, for which the question is not entirely settled, the ABJM amplitudes are known to receive contributions from both odd and even loops, i.e., 
\begin{align}
\mathbb{W}_n = \sum_{\ell = 0}^\infty g^{\ell} \mathbb{W}^{(\ell)}_n 
\, , 
\end{align}
where both $\mathbb{W}^{(\ell = {\rm even})}_n$ and $\mathbb{W}^{(\ell = {\rm odd})}_n$ are non-vanishing. In fact, it was demonstrated by an explicit calculation \cite{Bianchi:2012cq,Bargheer:2012cp,Brandhuber:2012wy,Brandhuber:2012un} that all one-loop amplitudes are proportional to the shifted tree amplitudes, 
\begin{align}
\mathbb{W}^{(1)} = \frac{\pi}{2} \mathbb{W}^{(0)}_{\rm shifted} \, ,
\end{align}
up to an overall step-function of kinematical variables, with $\mathbb{W}^{(0)}_{\rm shifted}$ $\sim \vev{\Phi_1\bar\Psi_2 \Phi_3  \dots \bar\Psi_{n}}_{\rm tree}$.%
\footnote{The step function vanishes for $n = 4$ and the one-loop four-leg amplitude is identically zero \cite{Agarwal:2008pu}.} They are thus rational functions. 
The two-loop amplitudes are functions of transcendentality two \cite{Chen:2011vv,Bianchi:2011dg,CaronHuot:2012hr}. This is consistent with would-be dual conformal anomaly 
equations which would predict the presence of the BDS function accompanied by the cusp anomalous dimension in addition to a remainder function of the conformal cross ratios. This was 
verified by a two-loop calculation of the six-leg amplitude in \cite{CaronHuot:2012hr}. Our focus in the subsequent discussion will be on the even part of the six-leg (hexagon) amplitude, leaving the flux-tube interpretation of the odd part to a future investigation.

\subsection{Hexagon data}

In order to carry out a systematic OPE analysis of the hexagon amplitude, we need to cast it in a right form and express it in terms of momentum twistors $Z_i$ and 
associated Grassmann variables $\eta_{i}$, with $i = 1, ..., 6$ enumerating the legs. At tree level, we can use a Yangian invariant form, that was derived in \cite{Gang:2010gy}, 
and latter recast in terms of momentum twistors in \cite{Elvang:2014fja}. It is given by the sum of two Yangian invariants $\mathcal{Y}_{1,2}$,
\begin{align}
\mathbb{W}_6^{(0)} = \mathcal{J} (\mathcal{Y}_1 + \mathcal{Y}_2)\, ,
\end{align}
which correspond to the $s = \pm$ terms in $I_2$, respectively, in Eq.\ (5.53) of \cite{Elvang:2014fja}. It is accompanied by a Jacobian $\mathcal{J}$, whose form we will recall shortly. 
The other linearly-independent combination of $\mathcal{Y}$'s determines the one-loop amplitude, which is expressed via the shifted tree amplitude by one site, i.e., 
\begin{align}
\mathbb{W}_{6, \rm shifted}^{(0)} = \mathcal{J} (\mathcal{Y}_1 - \mathcal{Y}_2)
\, .
\end{align}
Both the $\mathcal{Y}$'s and $\mathcal{J}$ are given in terms of the momentum twistors of the six-leg amplitude. For application to the collinear limit, we shall parameterise them in a conventional 
way, see Appendix A of Ref.\ \cite{Basso:2013aha}, using 
\beq\label{hex-twistors}
\left(\begin{array}{c} Z_{1}\\ Z_{2}\\ Z_{3}\\ Z_{4}\\ Z_{5}\\ Z_{6}\end{array}\right) 
= 
\left(
\begin{array}{cccc} 
e^{\sigma-\frac{i}{2}\phi}& 0 & e^{\tau+\frac{i}{2}\phi} & e^{-\tau+\frac{i}{2}\phi} \\ 
1& 0 & 0 & 0\\ 
-1& 0 & 0 & 1\\ 
0& 1 & -1 & 1\\ 0& 1& 0& 0\\ 0& e^{-\sigma-\frac{i}{2}\phi}& e^{\tau+\frac{i}{2}\phi}& 0 
\end{array}
\right)
\, ,
\eeq
with $\sigma, \tau,\phi$ being the 4d OPE coordinates introduced earlier. The reduction to 3d is obtained by imposing sp$(4)\sim$ so$(2,3)$ constraints on the twistors 
\cite{Elvang:2014fja}, namely,
\beq
\mathcal{hh}i, i+1\mathcal{ii} = 0\, , \qquad \forall i\, ,
\eeq
where the bracket is a symplectic form,
\beq
\mathcal{hh}i, j\mathcal{ii} = \Omega_{AB} Z^{A}_{i}Z^{B}_{j}\, , \qquad \Omega_{AB} = -\Omega_{BA}\, .
\eeq
Imposing these constraints on the hexagon twistors (\ref{hex-twistors}) enforces $e^{2i\phi} = 1$ and fixes
\beq\label{omega}
\Omega = \left(\begin{array}{cccc} 0& +1& 0& 0\\ -1& 0& 0& 0\\ 0& 0& 0& +1\\ 0& 0& -1& 0\end{array}\right)\, ,
\eeq
up to an overall factor. One can then plug the above twistors, and the double-angle brackets constrained in this manner, in the Grasmannian formulae and expand the amplitudes in the collinear 
limit $\tau \rightarrow\infty$. (As we stated above, we work with $\phi = 0$ parametrization in the following.) In particular, the Jacobian takes a very concise form%
\footnote{$\mathcal{J}$ differs slightly from the Jacobian $J_{234}$ given in Eq.~(5.38) of Ref.~\cite{Elvang:2014fja}, since we stripped out the Parke-Taylor prefactor $\sqrt{-\left<12\right>\left<23\right>...\left<61\right>}$ from the amplitude, see Eq.~(\ref{SuperAmplitudeAn}).}
\beq
\mathcal{J} = \sqrt{-\frac{\mathcal{hh}6, 2\mathcal{ii}\mathcal{hh}4, 6\mathcal{ii}\mathcal{hh}5, 1\mathcal{ii}^2}{\mathcal{hh}2, 4\mathcal{ii}}} = e^{\frac{1}{2}(\sigma+\tau)}\, .
\eeq
The fractional twist, that this factor implies, is essential for the proper flux-tube interpretation of the scattering amplitudes.

Besides constraining the kinematics, we also want to fix the R charge and select `good' components of the superamplitudes from the point of view of the OPE. Since the underlying $R$-symmetry 
is SU$(4)$ rather than SU$(3)$ (which is manifest), there are multiple relations among Grassmann components of the superamplitude \re{SuperAmplitudeAn}. In fact, there are only two 
nontrivial amplitudes that we have to extract. We choose them to be the coefficients in front of $\eta_1^3$ and $\eta_4^3$,
\begin{align}
\mathbb{W}_6 
= 
\eta_1^3 \mathcal{W}_{\psi} 
+
\eta_4^3 \mathcal{W}_{\phi} 
+ \dots
\, .
\end{align}
The reason is that this choice was natural in the SYM, where these amounted to replacing the incoming and outgoing vacua in the pentagon decomposition by charged vacua. We expect 
something similar here.

Plugging the constrained twistors in the formulae of Ref.\ \cite{Elvang:2014fja}, we get the two amplitudes
\beq\label{trees}
\begin{aligned}
&\mathcal{W}_{\phi}  = \mathcal{J} \times \frac{e^{-\tau}(e^{\sigma}+2e^{-\tau})}{(1+e^{-2\tau})(1+e^{2\sigma}+2e^{\sigma-\tau}+e^{-2\tau})}\, ,\\
&\mathcal{W}_{\psi}  = \mathcal{J} \times \frac{e^{-\tau}(1-e^{\sigma-\tau}-e^{-2\tau})}{(1+e^{-2\tau})(1+e^{2\sigma}+2e^{\sigma-\tau}+e^{-2\tau})}\, .
\end{aligned}
\eeq
Remarkably, these expressions coincide, up to the Jacobian, with the scalar and fermion components of the 6-leg SYM amplitude, at $\phi = 0$,
\beq
\mathcal{W}_{\phi} = \mathcal{J}\times \mathcal{W}^{(1144)}_{\mathcal{N}=4} \, , \qquad \mathcal{W}_{\psi} = \mathcal{J} \times \mathcal{W}^{(1444)}_{\mathcal{N}=4}\, ,
\eeq
hence we dressed them with the `boson' $\phi$ and `fermion' $\psi$ subscript, respectively. We measure now the importance of the Jacobian, it is adjusting the twists of what is flowing in 
the OPE channel. In the SYM theory all excitations have integer twists. Thanks to the Jacobian, in the ABJM theory, all excitations that are being exchanged have half-integer twists, 
implying that what flows has the quantum number of a spinon.

\subsection{Tree level OPE}

Let us proceed with the large $\tau$-expansion of the tree amplitudes. The leading-twist contributions at tree level are immediately found to be 
\begin{align}
\label{LOantispinonExchange}
\mathcal{W}_{\phi} 
= \e^{-\tau/2} \frac{\e^{3\sigma/2}}{1 + \e^{2\sigma}} + \dots\, , \qquad \mathcal{W}_{\psi} 
= \e^{-\tau/2} \frac{\e^{\sigma/2}}{1 + \e^{2\sigma}} + \dots\, .
\end{align}
They exhibit a clear signature of the exchanged particle to possess twist 1/2. It, therefore, must be a single spinon. We can thus propose a flux-tube representation in the form of a single 
integral over the momentum of the spinon,
\beq
\mathcal{W}_{\phi/\psi} = \int \frac{du}{2 \pi} 
\e^{ip_{Z}(u)\sigma-E_{Z}(u)\tau} \nu_{\phi/\psi} (u) + \dots\, .
\eeq
The weights $\nu_{\phi/\psi} (u)$ for production/absorption of the spinon can immediately be read off from the above expressions at tree level by an inverse Fourier transformation yielding
\beq
\nu_{\phi} (u) = \nu_{\psi}(-u) =  \frac{1}{2} \Gamma \left(\ft14 + \ft{i}{2} u \right) \Gamma \left(\ft34 - \ft{i}{2} u \right) = \frac{\pi}{2\cosh{\left(\frac{\pi}{2}(u+\frac{i}{2}) \right)}}\, .
\eeq
They are different from the measure (\ref{Mansatz}) that we had obtained earlier, and this is the case for a good reason. The latter measure has bad square root singularities and thus cannot 
be the image of a tree level amplitude. However, we note that we can view these weights as the measure dressed with the smearing factors introduced earlier to describe the insertions of 
hypermultiplets along the bosonic Wilson loop. Namely,
\beq
\nu_{\phi}(u)  \sim \mathcal{N}_{\psi}(u) \mu_{Z}(u) \mathcal{N}_{\phi}^*(u)\, .
\eeq
It is very suggestive that the spinon that is flowing on the `loop' $\mathcal{W}_{\phi}$ is produced as a fermion $\psi$ at the bottom and annihilated as a scalar $\phi$ at the top, --- and 
inversely for the $\mathcal{W}_{\psi}$. This hybrid nature is apparently needed to get a proper `propagator' with the singularity of a tree level amplitude. In comparison, the non-hybrid process 
$\mathcal{N}_{s}(u)\mu_{Z}(u)\mathcal{N}^*_{s}(u) \sim \Gamma(s+\frac{iu}{2})\Gamma(s-\frac{iu}{2})$ with $s = 1/4$ and $s=3/4$ for boson and fermion, respectively, has square root 
singularities in position space, since it is a Fourier transform of a free field propagator $\sim (\cosh{\sigma})^{-2s}$ for a field with the conformal spin $s$. (This relation is the square limit of 
equation (\ref{formula-s}) obtained by sending $\sigma_{1,2} \rightarrow -\infty$ and $\sigma = \sigma_{2}-\sigma_{1}$ held fixed.)

Let us finally note that the smearing factors cancel out in the product
\beq\label{SpinonMeasure}
\nu_{\phi}(u)\nu_{\psi}(u) = \frac{\sqrt{2\pi}}{2g}\mu_{Z}^2(u) + o(1) \, ,
\eeq
which, therefore, appears closely related to the spinon measure (\ref{Mansatz}), and hence to the scalar measure of the SYM theory, 
\beq
\mu_{\Phi}(u) = \frac{\pi g}{\cosh{(\pi u)}} + O(g^3)\, .
\eeq
Accordingly the effective measures $\nu_{\phi,\psi}$ can be seen as an alternative way of splitting the SYM scalar measure into two meromorphic factors.

Equipped with the weights for the fundamental spinons, we can try to make sense of the higher twist corrections. High-twist means higher particle number and particle production is 
generically suppressed at weak coupling. The only known exception is when the particles are being produced as small fermions. These are known to be the only extra particles needed for 
scattering amplitudes in the SYM theory through one loop, see the loop counting rules and discussion in Refs.\ \cite{Belitsky:2014sla,Belitsky:2014lta,Basso:2015rta,Cordova:2016woh,Lam:2016rel}. 
We expect the same to happen in the 3d theory meaning that all the higher twist corrections should arise from multiparticle states involving one spinon and an arbitrarily many small fermions 
attached to it, that is,
\beq\label{Hilbert}
\sum_{\textrm{states}} = \sum_{n = 0} Z\Psi^{2n} \oplus \bar{Z}\Psi^{2n+1}\, .
\eeq
An estimate of the weights of genuine multiparticle states suggests that (\ref{Hilbert}) is valid through two loops.%
\footnote{The estimate follows from considering the available twist $3/2$ states: $ZF, \bar{Z}\Psi, Z^2\bar{Z}, \bar{Z}^3$, with $\Psi$ a large fermion. For the corresponding integrands, we 
expect, schematically, 
\beq\notag
\mu_{ZF} \sim \frac{\nu_{\phi}\mu_{F}}{P_{Z|F}P_{F|Z}} \sim \mu_{\bar{Z}\Psi} \sim \frac{\nu_{\psi}\mu_{\Psi}}{P_{\bar{Z}|\Psi}P_{\Psi|\bar{Z}}}\sim g^4, \,\,\, 
\mu_{Z^2\bar{Z}} \sim \frac{\nu_{\phi}^2\nu_{\psi}}{(P_{Z|\bar{Z}}P_{\bar{Z}|Z})^2P_{Z|Z}^2}\sim \mu_{\bar{Z}^3} \sim \frac{\nu_{\psi}^3}{(P_{Z|Z})^6}\sim g^3\, .
\eeq
}
In the following, we demonstrate that the exact kinematical dependence of the tree amplitude can be recovered from the flux tube series (\ref{Hilbert}). In the next subsection, we verify that it is 
still so at two loops.

Let us address the first subleading term in order to demonstrate the structure and then generalize to an arbitrary number of small fermions. Take the $\phi$ component. The higher twist excitation 
has twist 
$3/2$ and arises from a single small fermion forming a string with a parent spinon $\bar{Z}$. We expect the integrand for the process to be given by
\beq\label{intpsiPsi}
\frac{i\nu_{\psi}(u) \mu_{\Psi}(\check{v})}{((u-v)^2+9/4)P_{Z|\Psi}(u|\check{v})P_{\Psi|Z}(\check{v}|u)}\, ,
\eeq
where we choose the $\psi$ weight for the spinon $\bar{Z}$.%
\footnote{As said earlier we do not have much control on global phase factors. We put an $i$ by hand in (\ref{intpsiPsi}) because it is needed for matching the tree amplitude. This factor might 
in principle be absorbed in a rescaling of the $Z\Psi$ transition.} This choice is natural in light of the `bosonic' nature of the component. The rational part is the matrix part for the projection 
$\bar{\textbf{4}}\otimes \textbf{6}\rightarrow \textbf{4}$.
The integration over the fermion boils down to picking up the residue $v = u-3i/2$. Using the formula (\ref{inZPsi}) for the transition between a small fermion and a spinon, we get the integrand 
for the twist-3/2 descendent of the spinon
\beq
i(u-3i/2)\nu_{\psi}(u)
\eeq
A similar argument would apply to the $\psi$ component, choosing this time $-i\nu_{\phi}(u)$ for the measure of the parent spinon. One verifies that the effective measures so obtained match perfectly with the next-to-leading term in the tree amplitudes,
\begin{align}
\mathcal{W}_{\phi} 
= \dots + \e^{-3 \tau/2} \frac{2\e^{\sigma/2}}{(1 + \e^{2 \sigma})^2} + \dots
\, , \qquad
\mathcal{W}_{\psi} 
=  \dots - \e^{-3\tau/2} \frac{\e^{3\sigma/2} (3 + \e^{2 \sigma})}{(1 + \e^{2 \sigma})^2} + \dots
\, .
\end{align}

\begin{figure}
\begin{center}
\includegraphics[scale=0.40]{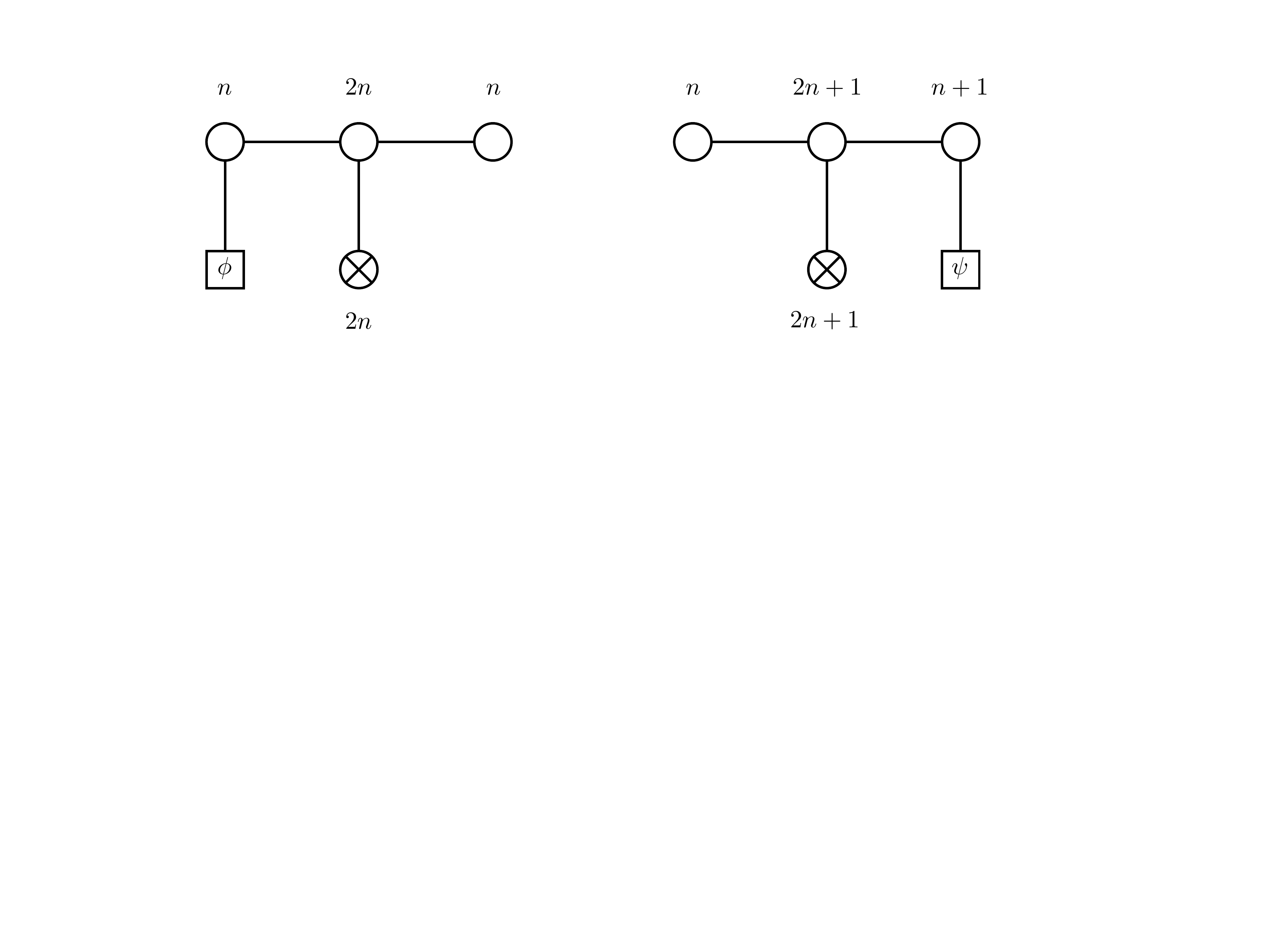}
\end{center}
\vspace{-7.5cm}
\caption{Dynkin diagrams encoding the matrix integral for descendents of a spinon $Z\sim \phi$ and anti-spinon $\bar{Z}\sim \psi$. The top part is the Dynkin diagram for the $SU(4)$ degrees of freedom with the labels indicating the numbers of corresponding auxiliary roots. They couple to the matter content represented by the bottom part. The small fermions are represented by the crossed nodes and the spinons by the boxes. The excitations numbers are chosen such that the overall charge matches the quantum number of $\phi$. We can view the small fermions as associated to a fermionic generator and extending the symmetry to OSp$(4|2)$.}\label{dm}
\end{figure}

We can generalize this story to strings of an arbitrary length, by carrying out the integral over the phase space of $n$ small fermions coupled to a spinon. Focusing on the $\phi$ component, the all-twist 
flux-tube representation that we put to the test is
\beq\label{all-twist}
\begin{aligned}
\mathcal{W}_{\phi} &= \sum_{n\geqslant 0} \int \frac{du}{2\pi} \nu_{\phi}(u) \int_{\mathcal{C}^{\circlearrowleft}} \frac{d^{2n}\textbf{v}}{(2\pi)^{2n}} \frac{\mu_{\Psi}(\check{\textbf{v}})\Pi^{(2n)}(u, \textbf{v})e^{iP\sigma-E\tau}}{P_{Z|\Psi}(u|\check{\textbf{v}})P_{\Psi|Z}(\check{\textbf{v}}|u)P^{\neq}_{\Psi|\Psi}(\check{\textbf{v}}|\check{\textbf{v}})} \\
&\,\,\, +i\sum_{n\geqslant 0} \int \frac{du}{2\pi} \nu_{\psi}(u)\int_{\mathcal{C}^{\circlearrowleft}} \frac{d^{2n+1}\textbf{v}}{(2\pi)^{2n+1}} \frac{\mu_{\Psi}(\check{\textbf{v}})\Pi^{(2n+1)}(u, \textbf{v})e^{iP\sigma-E\tau}}{P_{Z|\Psi}(u|\check{\textbf{v}})P_{\Psi|Z}(\check{\textbf{v}}|u)P^{\neq}_{\Psi|\Psi}(\check{\textbf{v}}|\check{\textbf{v}})} \, ,
\end{aligned}
\eeq
where $\textbf{v}$ denotes the set of fermion rapidities, with $2n$ and $2n+1$ elements in the first and second line, respectively, with $E, P$ being the total energy and momentum,
\beq\label{totalEP}
E = E_{Z}(u)+\sum_{i} E_{\Psi}(\check{v}_{i})\, , \qquad P = p_{Z}(u)+\sum_{i} p_{\Psi}(\check{v}_{i})\, ,
\eeq
and where, to save space, we introduced notations for functions of sets,
\beq
f(\textbf{w}) = \prod_{i}f(w_{i})\, , \qquad f^{\neq}(\textbf{w}, \textbf{w}) = \prod_{i\neq j} f(w_{i}, w_{j})\, .
\eeq
The small fermion contour $\mathcal{C}^{\circlearrowleft}$ goes anti-clockwise around all singularities in the lower half plane and $\Pi^{(k)}(u, \textbf{v})$ denotes corresponding matrix parts. 
The latter are bulky rational functions of rapidity differences, which can be written explicitly using formulae in \cite{Belitsky:2016vyq} or implicitly as a matrix-model-like integral over a set 
of SU$(4)$ auxiliary rapidities \cite{Basso:2015uxa}. The simplification that comes about here is that the small fermions can be understood as extending the latter matrix-model integral into 
the one for a system with OSp$(4|2)$ symmetry. Namely, using the weak coupling expressions for the small fermion transition and measure yields the integral
\beq\label{OSP}
\Pi^{(k)}_{{\rm OSp}(4|2)}(u) = \int_{\mathcal{C}^{\circlearrowright}} \frac{d^{k}\textbf{v}}{k!(2\pi)^k} h(\textbf{v}) \prod_{i<j}(v_{i}-v_{j})^2 \Pi^{(k)}(u, \textbf{v})\, ,
\eeq
where
\beq
h(\textbf{v}) = \frac{(-1)^k\mu_{\Psi}(\check{\textbf{v}})}{P_{Z|\Psi}(u|\check{\textbf{v}})P_{\Psi|Z}(\check{\textbf{v}}|u)P_{\Psi|\Psi}^{\neq}(\check{\textbf{v}}|\check{\textbf{v}})\prod_{i< j}(v_{i}-v_{j})^2} 
= \prod_{i} v_{i} \times (1+O(g^2)) 
\eeq
is a symmetric function of the fermion rapidities. Importantly, the self-interaction of the fermions reduces to a Vandermonde determinant, as expected for a fermionic node. Combining this integral 
with the integral representation for $\Pi^{(k)}(u, \textbf{v})$, see rules in \cite{Basso:2015uxa}, one immediately identifies in the pattern of the couplings the Dynkin diagram of OSp$(4|2)$, as 
pictured in figure \ref{dm}.

\begin{figure}
\begin{center}
\includegraphics[scale=0.40]{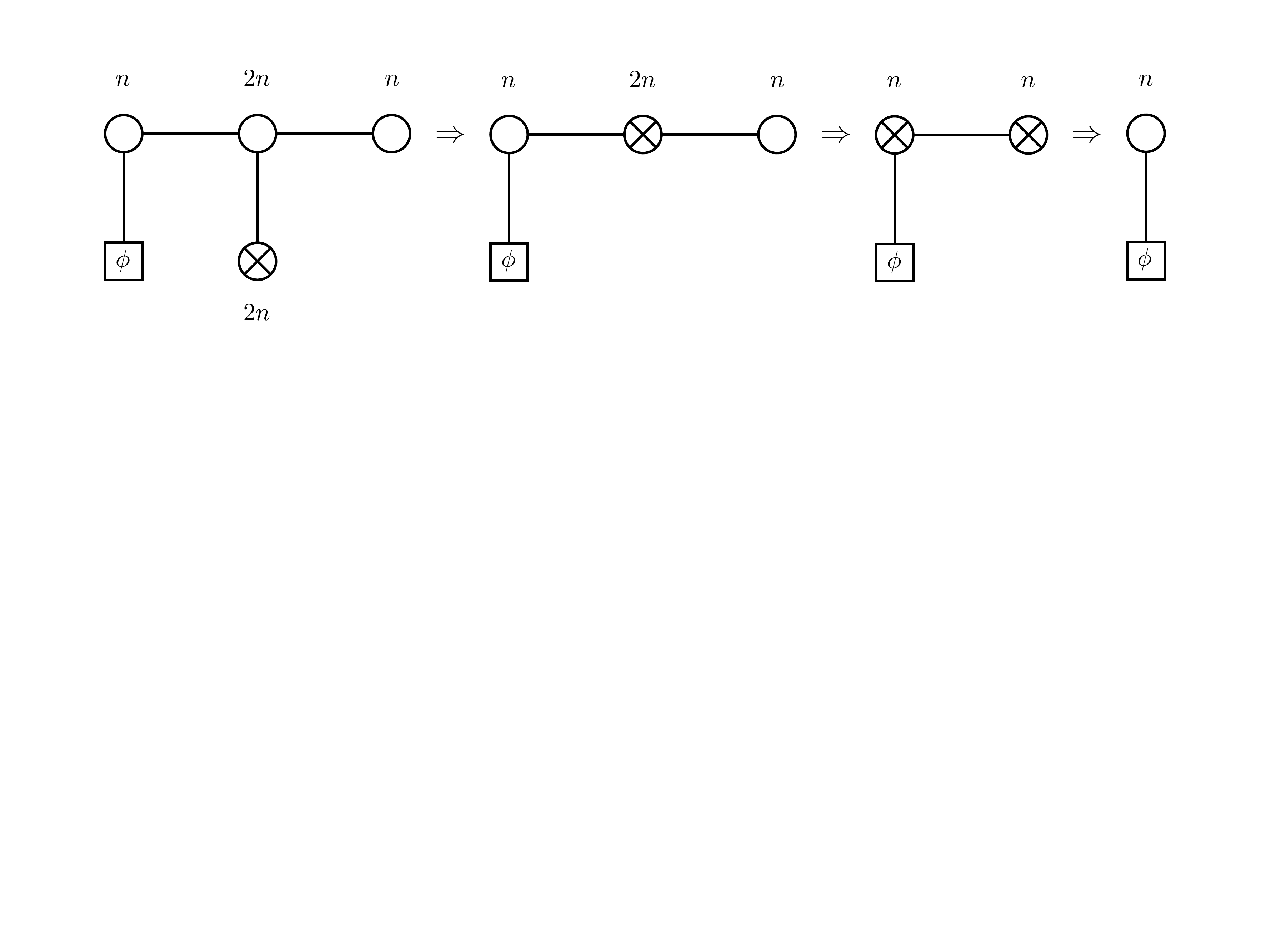}
\end{center}
\vspace{-7.5cm}
\caption{Integrating out the small fermions yields an effective matrix integral for a new system with the matrix structure depicted here. One can keep going until one is left with a single node 
coupled to the spinon. The structure of the latter node is akin to the one for an SL$(2)$ spin chain. The manipulations carried out here are reminiscent of the dualization procedure for super 
spin chains.}\label{dd}
\end{figure}

One can then integrate out the nested integrals in (\ref{OSP}) starting with the fermions. The excitation numbers, shown in figure \ref{dm}, are indicating the number of integration variables per 
node and are such that a unique pattern of residues is allowed at every step. E.g., at the first step, the $k$ fermions rapidities must bind below the $k$ auxiliary roots $\textbf{w}$ they couple to,
\beq
\int_{\mathcal{C}^{\circlearrowright}} \frac{d^{k}\textbf{v}}{k!(2\pi)^k} h(\textbf{v}) \times \frac{\prod_{i<j}(v_{i}-v_{j})^2}{\prod_{i, j}((v_{i}-w_{j})^2+ 1/4)} = h(\textbf{w}^{[-1]})\times \prod_{i\neq j} \frac{1}{(w_{i}-w_{j})^2+1}\, ,
\eeq
where $\textbf{w}^{[-a]} = \{w_{i}-ia/2\}$. The right-hand-side cancels a similar factor present in the self-interaction of the SU$(2)$ rapidities $\textbf{w}$ which are then left to interact by means of 
a Vandermonde determinant. Said differently, the rapidities $\textbf{w}$ are fermionised and one can iterate the procedure. The steps are schematised in figure \ref{dd} and mimic the dualisation 
of nested Bethe ansatz equations for super-spin chains. At the end of the process, one is left with an effective integral for an SL$(2)$ system,
\beq
\begin{aligned}
\Pi^{(2n)}_{{\rm OSp}(4|2)}(u) &= \int_{\mathcal{C}^{\circlearrowright}} \frac{d^{n}\textbf{w}}{2^{n}n!(2\pi)^{n}} h(\textbf{w}^{[-2]} \cup \textbf{w}^{[-4]})\prod_{i}\frac{1}{((u-w_{i})^2+1/4)} \prod_{i<j} \frac{(w_{i}-w_{j})^2}{(w_{i}-w_{j})^2+4} \\
&= \frac{1}{(2n)!} \prod_{j=1}^{2n}h(u^{[-1-2j]}) \, ,
\end{aligned}
\eeq
and similar one for the odd cases.

The punch line is that only 1 string remains given a $k$, namely, a length $k$ string attached below the spinon at a distance $3i/2$. This is quite remarkable given that the fermions here are in 
the vector representation, which offers a wider patterns of strings a priori. E.g., the two fermion integral discussed in Appendix \ref{App:fermions} produces two type of strings that both contribute 
in the end. A similar pattern of strings, although more complicated, was found in the higher twist analysis of the tree and loop amplitudes in the SYM theory \cite{Cordova:2016woh,Lam:2016rel}. 
On the field theory side, we can view the length $k$ string as describing the twist $k+1/2$ descendent $\mathcal{D}_{12}^k \phi$. The answer is simpler than in the 4d case since we do 
not need to include powers of $\mathcal{D}_{22} \sim \partial_{\tau}$ in the OPE. Owing to the equation of motion $\mathcal{D}_{11}\mathcal{D}_{22}\phi \sim \mathcal{D}_{12}^2\phi$ and thus in 
the large spin background the twist 2 derivative $\mathcal{D}_{22}$ can be traded for $\mathcal{D}_{12}^2$. (A similar argument works for the fermion.) Hence only one type of strings is needed 
to span all the field-excitations.

In the end, once all strings are formed, we obtain the flux tube representation of the tree level amplitudes,
\begin{align}\label{promote}
\mathcal{W}_{\phi/\psi} = \e^{-\tau/2} \sum_{n=0}^\infty (-1)^n\e^{- 2 n \tau} \int \frac{du}{2 \pi} \e^{i u \sigma} 
\left[ \frac{ (i u + \ft32)_{2n}}{(2n)!} \nu_{\phi/\psi} (u) \pm \e^{- \tau} \frac{ (i u + \ft32)_{2n + 1}}{(2n + 1)!} \nu_{\psi/\phi} (u) \right]
\, ,
\end{align}
where the trace of the small fermions is encoded in the Pochhammer symbol
\beq
(\alpha)_n = \Gamma(\alpha+n)/\Gamma(\alpha)\, .
\eeq
One easily verifies that the above series matches with the higher twist terms in (\ref{trees}). One can actually do better and resums the OPE, in the spirit of what was done in the SYM theory \cite{Cordova:2016woh,Lam:2016rel}. All one needs to note is the relation
\begin{align}
\label{OPEresum1}
\sum_{n = 0}^\infty(-1)^n \frac{\left(i u + \ft32 \right)_{2n}}{(2n)!}   \e^{- 2 n \tau} = \frac{1}{2} \left[(1 + i\e^{-\tau})^{- 3/2 - i u} +(1 - i\e^{-\tau})^{- 3/2 - i u} \right]
\, , 
\end{align}
and an analogous one for odd $n$'s, which merely yields a sign change in front of the second term in brackets and an overall factor of $i$. With their helps, we can write the flux tube series (\ref{promote}) as
\beq\label{promote2}
\begin{aligned}
\mathcal{W}_{\phi/\psi} = \frac{1}{2}\bigg[\frac{\nu_{\phi/\psi}(\sigma_{+})}{(1+ie^{-\tau})^{3/2}}+\frac{\nu_{\phi/\psi}(\sigma_{-})}{(1-ie^{-\tau})^{3/2}}\bigg] \pm \frac{i}{2}\bigg[\frac{\nu_{\psi/\phi}(\sigma_{+})}{(1+ie^{-\tau})^{3/2}}-\frac{\nu_{\psi/\phi}(\sigma_{-})}{(1-ie^{-\tau})^{3/2}}\bigg]\, ,
\end{aligned}
\eeq
where $\nu_{\phi}(\sigma)$ and $\nu_{\psi}(\sigma) = \nu_{\phi}(-\sigma)$ are the twist 1/2 seeds \re{LOantispinonExchange} and where
\beq
\sigma_{\pm} = \sigma -\log{(1\pm i e^{-\tau})}\, .
\eeq
One easily verifies that these expressions agree with the tree amplitudes (\ref{trees}) at any $\tau$.

\subsection{Loop level OPE}

After this initial success, let us move on to the two-loop analysis of $\mathbb{W}_6$. The two loop ratio function $\mathbb{R}_6^{(2)}$ was computed in \cite{CaronHuot:2012hr}, under the 
assumption of cut-constructibility  of the amplitude from a set of dual-conformal invariant integrals, and was cast in the form of the tree amplitudes dressed by transcendentality-two functions 
of the conformal cross ratios $u_j$,
\begin{align}
\label{P6}
\mathbb{R}_6^{(2)} & = \ft12 \mathbb{W}^{(0)}_6 \sum_{i = 1}^3 \left[ - 2\pi^2 + {\rm Li}_2 (1 - u_i) + \ft12 \log u_i \log u_{i+1}+ ({\arccos} \sqrt{u_i})^2 \right] \nonumber\\
&\,\, + \ft{1}{2} \mathbb{W}^{(0)}_{6, \textrm{shifted}} \sum_{i=1}^3 {\arccos} \, \sqrt{u_i} \log \frac{u_{i + 1}}{u_{i+2}} \, ,
\end{align}
with implied cyclicity $u_{j+3} = u_j$. Notice that we eliminated the BDS contribution from the result of \cite{CaronHuot:2012hr} according to the definition \re{RatioFunctionPn}. Translated to our language, it means that
\begin{align}
\label{W2}
\mathbb{W}^{(2)}_6 = \mathbb{R}^{(2)}_6 + \frac{1}{2}\mathbb{W}^{(0)}_6 r_6  
\, ,
\end{align}
according to \re{WLobservable} and (\ref{W6}). To evaluate it we need the shifted tree amplitudes. They happen to be directly related to our component amplitudes and read
\beq
\mathcal{W}_{\phi/\psi, \textrm{shifted}} = \eta_{\phi/\psi} \mathcal{W}_{\psi/\phi}\, ,
\eeq
up to a sign $\eta_{\phi/\psi} = -/+$ .

We can then expand the two loop formula at leading twist, using the parameterization (\ref{OPEpar}) for the hexagon cross ratios, and obtain
\begin{align}
\label{W2phi}
\mathcal{W}^{(2)}_{\phi}
& =  \tau \e^{- \tau/2} \frac{\e^{3\sigma/2}}{1 + \e^{2 \sigma}} \left[\log{(e^{\sigma} + e^{-\sigma})} -ie^{-\sigma}\log{\left(\frac{e^\sigma + i}{e^\sigma - i}\right)} - \tfrac{1}{2}\pi e^{-\sigma} \right] \\
&+ \e^{- \tau/2} \frac{\e^{3\sigma/2}}{1 + \e^{2 \sigma}} \bigg[-\tfrac{7}{12}\pi^2 + \tfrac{1}{4}\log{\left(\frac{e^{\sigma}+i}{e^{\sigma}-i}\right)}\log{\left(\frac{e^{-\sigma}+i}{e^{-\sigma}-i}\right)}-\tfrac{1}{4}\log{(e^{2\sigma}+1)}\log{(e^{-2\sigma}+1)} \nonumber\\
&\,\,\,\,\, \qquad\qquad\qquad\,\,\,\,\,\,\,\, +  \tfrac{1}{2}ie^{-\sigma} \log{\left(\frac{e^{\sigma}+i}{e^{\sigma}-i}\right)}\log{(e^{\sigma}+e^{-\sigma})} +\tfrac{1}{4}\pi e^{-\sigma}\log{(e^{2\sigma}+1)}\bigg] + \dots \, , \nonumber
\\
\label{W2psi}
\mathcal{W}^{(2)}_{\psi}
& =  \tau \e^{- \tau/2}  \frac{\e^{\sigma/2}}{1 + \e^{2 \sigma}} \left[\log{(e^{\sigma} + e^{-\sigma})} + ie^{\sigma} \log{\left(\frac{e^\sigma + i}{e^\sigma - i}\right)} + \tfrac{1}{2}\pi e^{\sigma}  \right] \\
&+  \e^{- \tau/2}  \frac{\e^{\sigma/2}}{1 + \e^{2 \sigma}} \bigg[-\tfrac{7}{12}\pi^2+\tfrac{1}{4}\log{\left(\frac{e^{\sigma}+i}{e^{\sigma}-i}\right)}\log{\left(\frac{e^{-\sigma}+i}{e^{-\sigma}-i}\right)}-\tfrac{1}{4}\log{(e^{2\sigma}+1)}\log{(e^{-2\sigma}+1)}\nonumber\\
&\,\,\,\,\, \qquad\qquad\qquad\,\,\,\,\,\,\,\, -  \tfrac{1}{2}ie^{\sigma} \log{\left(\frac{e^{\sigma}+i}{e^{\sigma}-i}\right)}\log{(e^{\sigma}+e^{-\sigma})} -\tfrac{1}{4}\pi e^{\sigma}\log{(e^{2\sigma}+1)}\bigg] + \dots \, , \nonumber
\end{align}
which are such that $\mathcal{W}_{\psi}(\sigma) = \mathcal{W}_{\phi}(-\sigma) +O(e^{-3\tau/2})$. We immediately observe that these expressions contain $\tau$-enhanced terms. These are leading discontinuities, which according to (\ref{LOantispinonExchange}) should arise from the expansion of the spinon energy to the first order in $g^2$,
\begin{align}
-\tau\e^{-\tau/2} \int \frac{du}{2 \pi} \e^{i u\sigma } \nu^{\textrm{tree}}_{\phi} (u) E^{(2)}_{Z} (u) \, .
\end{align}
This flux-tube integral is easily verified to reproduce the first lines in (\ref{W2phi}) and (\ref{W2psi}) using the expression (\ref{EZ1}) for the two loop spinon energy $E^{(2)}_{Z}$.

The remaining terms in Eqs.\ \re{W2phi} and \re{W2psi} have a number of origins. Some of them stem from the correction $p^{(2)}_{Z}(u)$ to the spinon momentum, see (\ref{EZ1}), and 
some from the correction to the spinon weights,
\beq
\nu_{\phi/\psi}(u) = \nu^{\textrm{tree}}_{\phi/\psi}(u) (1+O(g^2))\, .
\eeq
The two cases can be accommodated in the expression
\begin{align}
\e^{-\tau/2} \int \frac{du}{2 \pi} \e^{i u \sigma} 
\nu_{\phi/\psi}^{\textrm{tree}} (u) \left[ i p_Z^{(2)} (u) \sigma + \delta \mu (u) \right]
\, ,
\end{align}
with the same $\delta \mu$ for both the $\phi$- and $\psi$-components. Furthermore, the most complicated part of the shift in weights is given by half the shift of the scalar measure in SYM, namely,
\beq\label{two-loop-mu}
\delta \mu (u)  = \ft12 \delta \mu_\Phi (u) - \pi^2  {\rm sech}^2 (\pi u) - 2 \zeta_2 \, ,
\eeq
with \cite{Basso:2013aha}
\beq
\delta \mu_\Phi (u) = 8\zeta_{2}-2\pi^2{\rm sech}^2 (\pi u)-2H_{iu-\frac{1}{2} }H_{-iu-\frac{1}{2}} \, ,
\eeq
where $\zeta_{2} = \zeta(2) = \pi^2/6$ and $H_{z}=H(z) = \psi(1+z)-\psi(1)$.

The loop correction (\ref{two-loop-mu}) might come from the smearing factors $\mathcal{N}_{\phi,\psi}$ and/or the measure $\mu_{Z}$. In the latter case it would be the first perturbative 
evidence that our ans\"atze for the spinon pentagons must be corrected. E.g., if we assume that formula (\ref{SpinonMeasure}) is valid through two loops and discard possible odd loop 
effects then the first correction to $f'(u)$ in (\ref{muZTomuN4}) is fixed by (\ref{two-loop-mu}) to be
\beq
f'(u) = 1+ 2g^2 \left(\pi^2  {\rm sech}^2 (\pi u) + 2 \zeta_2\right) +\ldots\, .
\eeq
It can alternatively be written as a correction to $\alpha   = 1 +O(g^2)$ in (\ref{Wansatz}).

As done at tree level above, the knowledge of the lowest twist components opens a way for an all-twist resummation of the OPE at two loops with minor modifications.

Let us begin with the terms linear in $\tau$. Here we simply need to note that the small fermion energy (\ref{small-p}) is not corrected at $O(g^2)$. Hence, each term in (\ref{promote}) gets dressed 
with the same spinon energy $E_{Z}^{(2)}$, independently of the twist. We can therefore re-sum the OPE by plugging in (\ref{promote2}) the leading discontinuities at leading twist and verify that they match with the terms $\propto \tau \sim -\frac{1}{2}\log{u_{2}}$ in (\ref{W2}).

We can also test the all-twist OPE formula (\ref{all-twist}) for the term in $\tau^0$. All we need to do to accomplish this is to keep the first sub-leading term in the perturbative expansion of the 
$Z\Psi$ and $\Psi\Psi$
pentagons,
\beq\label{weight-shift}
\begin{aligned}
\frac{\mu_{\Psi}(\check{v})}{P_{Z|\Psi} (u|\check{v})P_{\Psi|Z} (\check{v}|u)}  &= -\left(v - \pi g^2\textrm{tanh}{(\pi u)}  + O (g^4) \right) \, , \\
\frac{1}{P_{\rm \Psi|\Psi} (\check{u}|\check{v})P_{\rm \Psi|\Psi} (\check{v}|\check{u})} & = (u - v)^2 \left(1 + 2g^2/uv + O (g^4) \right) \, ,
\end{aligned}
\eeq
and plug them into (\ref{OSP}). The first term shifts the weight of each fermion, while the second one slightly corrects the pairwise interaction between fermions. Putting everything 
together and taking the string pattern into account gives
\begin{align}
\mathcal{W}^{(2)}_{\phi/\psi}\, |_{\tau^0}
&
= \e^{- \tau/2} \sum_{n = 0}^\infty \frac{\e^{- n\tau}}{n!} \int \frac{d u}{2 \pi} \e^{i u \sigma} \nu^{(n)}_{\phi/\psi} (u) (iu +\ft{3}{2})_n \\
&\,\,\,  \times \bigg\{ \delta \mu(u) +  i \sigma p^{(2)}_Z (u) + i \sigma  \sum_{j=1}^n p^{(2)}_{\Psi} \left(\check{u} - i (\ft{1}{2} + j)\right) \nonumber\\
&\qquad\qquad\qquad
- 
\sum_{j=1}^n \frac{\pi \tanh (\pi u)}{u - i (\ft{1}{2} + j)}
+ \sum_{j > k = 1}^n \frac{2}{\left( u - i (\ft{1}{2} + j) \right) \left(u - i (\ft{1}{2}+ k) \right)}
\bigg\}
\, , \nonumber
\end{align}
where in the first line $\nu^{(2n)}_{\phi/\psi} = (-1)^n \nu_{\phi/\psi}$ and $\nu^{(2n+1)}_{\phi/\psi} = \pm (-1)^n \nu_{\psi/\phi}$. The second line contains the two loop correction to the 
spinon measure (\ref{two-loop-mu}) and the loop correction to the total momentum $P$, which comes from the spinon and the string of small fermions attached to it, see (\ref{totalEP}) 
and (\ref{small-p}). Finally, the third line contains the shifts (\ref{weight-shift}).

To perform the resummation is not more difficult than for trees. In addition
to Eq.\ \re{OPEresum1}, we merely need two more results
\begin{align}
&
\sum_{n = 0}^\infty \frac{ \left(i u + \ft32 \right)_{2n}}{(2n)!} \e^{- 2 n \tau}  \sum_{j = 1}^{2n} \frac{1}{u - i (\ft12 + j)} 
= 
\partial_u \left( \sum_{n = 0}^\infty \frac{\left(i u + \ft32\right)_{2n}}{(2n)!}   \e^{- 2 n \tau} \right)
\, , \\
&
\sum_{n = 0}^\infty \frac{ \left(i u + \ft32\right)_{2n}}{(2n)!} \e^{- 2 n \tau}  \sum_{j > k = 1}^{2n} \frac{2}{\left( u - i (\ft{1}{2} + j) \right) \left( u - i (\ft{1}{2}+ k) \right)}
= 
\partial^2_u \left( \sum_{n = 0}^\infty \frac{\left(i u + \ft32\right)_{2n}}{(2n)!}   \e^{- 2 n \tau} \right)
\, , \nonumber
\end{align}
and analogous ones for odd $n$ with corresponding changes. The Fourier transform in rapidity is performed by means of the known leading twist expression at two loop 
order \re{W2phi} and \re{W2psi}. The resulting two-loop expression coincides with the corresponding components of Eq.\ \re{W2} with \re{P6}.

Having reproduced the two-loop hexagon within the pentagon OPE, let us finish with a few predictions. Since presently we are unable to unambiguously find all-order
expressions for all building blocks of the ABJM pentagon program, we will limit ourselves to the four loop leading discontinuities $\propto g^4 \tau^2$,
\beq\label{LD4}
\mathcal{W}_{\phi/\psi}^{(4)} =  \tau^2 (\nu_{\phi/\psi}^{\tau^2}(\sigma) e^{-\tau/2} + O(e^{-3\tau/2}) ) + O(\tau)\, .
\eeq
They arise from the insertion of the second power of the spinon energy into the leading order flux-tube integrands. We find
\beq
\begin{aligned}
&\nu_{\phi}^{\tau^2}(\sigma) = \int \frac{d u}{2 \pi} \e^{i u \sigma} \nu^{\textrm{tree}}_\phi (u) ( E^{(2)}_Z (u))^2 \\
&\,\, = \frac{\e^{\sigma /2}}{\e^{2 \sigma }+1} \bigg[\tfrac{1}{2} \e^{\sigma} \sigma ^2+ \tfrac{3}{2} \pi  \sigma + \ft14 \zeta_2 \e^{\sigma} +\e^{\sigma} \log^2{\left(\frac{\e^{\sigma }+i}{\e^{\sigma }-i}\right)} +\e^{\sigma} \log ^2{(\e^{2\sigma }+1)}\\
&\,\,\,\, - (2 \e^{\sigma} \sigma +\pi) \log{(\e^{2 \sigma }+1)} + i \log{\left(\frac{\e^{\sigma }+i}{\e^{\sigma }-i}\right)} (2 \sigma -\pi  \e^{\sigma }-2 \log{(\e^{2 \sigma }+1)}) \bigg] \, ,
\end{aligned}
\eeq
and
\beq
\nu_{\psi}^{\tau^2}(\sigma) =\nu_{\phi}^{\tau^2}(-\sigma)\, .
\eeq
The formulae can be upgraded to higher twists, such as to produce all terms in brackets in (\ref{LD4}), by applying the recipe (\ref{promote2}) to $\nu^{\tau^2}_{\phi/\psi}(\sigma)$.

\section{Discussion}\label{Sec5}

With this work, we initiated a systematic application of the pentagon program to the $\mathcal{N} = 6$ supersymmetric Chern-Simons
theory with matter. Presently, we addressed pentagon transitions for all fundamental excitations propagating on the ABJM flux tube. While the twist-one fermions 
and gluons (as well as all bound states thereof) were fixed uniquely, the spinons could not be constrained in a complete fashion. However, it did not create an obstruction for 
the applications that we were interested in.

Namely, in this paper we made a small step towards implementing the pentagon program for ABJM amplitudes. A 
success of this bold endeavor was not warranted as, contrary to their SYM counterparts, the dual description in terms of Wilson loops is not known and a naive 
supersymmetrization of the latter did not provide an adequate dual description for amplitudes with more than four legs. The fact that we could use the 
pentagon paradigm for their description within the same framework provides some new evidence for the existence of an observable that unifies both the ABJM Wilson 
loop and amplitudes under the same umbrella. It is unclear at this moment what it is, though.

There are a number of avenues open for future considerations which, at the same time, will make our conclusions more precise. The one of paramount
importance is dedicated efforts in higher loop calculations of scattering amplitudes. In particular, a two-loop eight-leg analysis would provide explicit
data to constraint the spinon pentagons directly as a function of the two rapidities, rather than just one through the spinon measure, as we performed in this
study. This amplitude is within reach within the generalized unitarity framework since contributing graph topologies are the same as for the 
ABJM hexagon\footnote{We would like to thank Simon Caron-Huot for correspondence on this issue.}. Another very valuable piece of data would come from three-loop
hexagon since it would clarify the pentagon structure for the odd part of the amplitude.

Having these at our disposal would put the framework on a firmer foundation, as it would allow one to point the way for proper implementation of the mirror 
axiom for the spinon excitations. Hopefully future studies along these lines will endow ABJM amplitudes 
with a dual Wilson-loop-like observable and will, therefore, make the application of pentagons fully justified.

\section*{Acknowledgments}

We thank Simon Caron-Huot, Amit Sever and Pedro Vieira for discussions. The research of A.B.\ was supported by the U.S. National Science Foundation under the grant PHY-1713125. 
The research of B.B.\ was supported by the French National Agency for Research grant ANR-17-CE31-0001-02.

\appendix

\section{Fermions at strong coupling}\label{App:fermions}

In this appendix we discuss the two fermions integral at strong coupling and compare its prediction with the string theory answer for the mass 2 boson. We refer the reader to \cite{Basso:2014koa,Bonini:2015lfr,Bonini:2018mkg} for detailed analysis in the SYM theory. All we need to know is that the two fermion integrand (\ref{psipsi}) can be written as
\beq\label{int2f}
\frac{3(u_{1}-u_{2})^2 du_{1}du_{2}}{\left((u_{1}-u_{2})^2+1\right)\left((u_{1}-u_{2})^2+4\right)} \Sigma(u_{1}, u_{2})\, ,
\eeq
where $\Sigma(u_{1}, u_{2}) = 1+ O(1/g)$ at strong coupling in the regime of interest, $(2g)^2 < u_{1,2}^2$. Naively, after rescaling the rapidities $u_{1,2}\rightarrow 2g u_{1,2}$, the above integrand is of order $O(g^0)$ and thus should not enter the computation of the minimal area $\mathcal{A} = O(g)$. This is overlooking that the integration contours get pinched between the lower and upper half plane poles. Deforming the contours and picking the residues lead to single particle like contributions that are of the right order $O(g)$. In the case at hand we get two strings of fermions corresponding to the poles at $u_{1}-u_{2} = 2i$ and $u_{1}- u_{2} = i$ in (\ref{int2f}). These strings are degenerate at strong coupling and both behave like a mass 2 boson. The sum of their residues yields $\frac{1}{2}du$ for the measure of their center of mass. In comparison, the two fermion integral in the SYM has the structure
\beq
\frac{4 du_{1}du_{2}}{((u_{1}-u_{2})^2+4)} \Sigma'(u_{1}, u_{2})\, ,
\eeq
where $\Sigma' = 1+\ldots$ and thus offers a single string at $u_{1} - u_{2} = 2i$ with unit residue $du$. From there it follows that per unit of $g$ the 2 fermion contribution to the minimal area in $\textrm{AdS}_{4}$ is half the one for $\textrm{AdS}_{5}$, in agreement with the string theory prediction.

\bibliography{biblio}
\bibliographystyle{JHEP}

\end{document}